\documentclass[twocolumn]{aastex701}

\usepackage{url} 
\usepackage[T1]{fontenc}
\usepackage{amsmath,amssymb}
\usepackage{bm}
\usepackage{subcaption}
\usepackage{soul}
\usepackage{hhline}
\usepackage{comment}
\usepackage{natbib}
\usepackage[dvipsnames]{xcolor}
\usepackage{gensymb}
\usepackage{hyperref}
\usepackage{aas_macros}
\usepackage{graphicx}

\newcommand{\rs}{R_\odot}
\newcommand{\dash}{\text{--}}
\newcommand{\kmps}{\text{km}\,\text{s}^{-1}}

\begin{document}


\title{Probing the Outer Edge of the Sun's Alfv\'en Zone with Early-Mission PUNCH Observations --- Fourier Motion Filtering and Inbound Flows}

\author[orcid=0000-0002-7174-6948]{Rohit Chhiber} 
\affiliation{Department of Physics \& Astronomy, University of Delaware, Newark, DE 19716, USA}
\affiliation{Heliophysics Science Division, NASA Goddard Space Flight Center, Greenbelt, MD 20771, USA}
\email[show]{rohit.chhiber@nasa.gov}

\author[orcid=0000-0002-3699-3134]{Steven R. Cranmer} 
\affiliation{Department of Astrophysical and Planetary Sciences,
Laboratory for Atmospheric and Space Physics,
University of Colorado, Boulder, CO 80309, USA}
\email[show]{steven.cranmer@colorado.edu}

\author[orcid=0000-0002-7164-2786]{Craig E. DeForest}
\affiliation{Southwest Research Institute, Boulder, CO, USA}
\email[]{deforest@boulder.swri.edu}

\author[orcid=0000-0001-8504-2725]{Ritesh Patel}
\affiliation{Southwest Research Institute, Boulder, CO, USA}
\email[]{ritesh.patel@swri.org}

\author[orcid=0000-0001-9831-2640]{Sarah Gibson}
\affiliation{National Center for Atmospheric Research / High Altitude Observatory, Boulder, CO, USA}
\email{sgibson@ucar.edu}

\author[orcid=0000-0003-4312-6298]{Raphael Atti\'e}
\affiliation{George Mason University, Innovation Park, VA, USA}
\affiliation{Heliophysics Science Division, NASA Goddard Space Flight Center, Greenbelt, MD 20771, USA}
\email{rattie@gmu.edu}

\author[orcid=0000-0003-4168-590X]{Francesco Pecora}
\affiliation{Department of Physics \& Astronomy, University of Delaware, Newark, DE 19716, USA}
\email{fpecora@udel.edu}

\author[orcid=0000-0003-2409-3742]{Nour E. Raouafi}
\affiliation{Johns Hopkins University Applied Physics Laboratory, Columbia, MD, USA}
\email{nour.rawafi@jhuapl.edu }

\author[orcid=0000-0001-7224-6024]{William H. Matthaeus} 
\affiliation{Department of Physics \& Astronomy, University of Delaware, Newark, DE 19716, USA}
\email[]{whm.edu}

\author[orcid=0000-0003-2965-7906]{Yan Yang}
\affiliation{Department of Physics \& Astronomy, University of Delaware, Newark, DE 19716, USA}
\email{yanyang@udel.edu}

\author[orcid=0000-0003-0021-9056]{Chris Gilly}
\affiliation{North-West Research Associatest, Boulder, CO, USA}
\email{gilly@nwra.com }

\author[orcid=0000-0001-5066-8509]{Anna Malanushenko}
\affiliation{High Altitude Observatory / National Center for Atmospheric Research, Boulder, CO, USA}
\email{anny@ucar.edu}

\author[orcid=0000-0001-6978-9765]{Iver H. Cairns
}
\affiliation{University of Sydney, Sydney, Australia}
\email{iver.cairns@sydney.edu.au}

\author{Ronnie Killough}
\affiliation{Southwest Research Institute, San Antonio, TX, USA}
\email{}
\author[orcid=0000-0002-4507-3372]{Jillian Redfern}
\affiliation{Southwest Research Institute, Boulder, CO, USA}
\email{}
\author{William L. Wells}
\affiliation{Southwest Research Institute, San Antonio, TX, USA}
\email{}
\author[orcid=0009-0004-0646-0523]{Matt N. Beasley}
\affiliation{Southwest Research Institute, Boulder, CO, USA}
\email{}
\author[orcid=0009-0000-6578-326X]{Glenn T. Laurent}
\affiliation{Southwest Research Institute, Boulder, CO, USA}
\email{}
\author[orcid=0000-0002-3253-4205]{Robin C. Colaninno}
\affiliation{Naval Research Laboratory, Washington, DC, USA}
\email{}
\author[orcid=0000-0002-4170-7104]{Nick R. Waltham}
\affiliation{United Kingdom Research \& Innovation / Science \& Technology Facilities Council / Rutherford Appleton Laboratory, Harwell Campus, Oxfordshire, U.K.}
\email{}
\author[orcid=0000-0003-3410-7650]{J. Marcus Hughes}
\affiliation{Southwest Research Institute, Boulder, CO, USA}
\email{}
\author[orcid=0000-0002-0494-2025]{Daniel B. Seaton}
\affiliation{Southwest Research Institute, Boulder, CO, USA}
\email{}
\author[orcid=0000-0002-4472-8517]{Samuel Van Kooten}
\affiliation{Southwest Research Institute, Boulder, CO, USA}
\email{}

\author[orcid=0000-0001-8318-8229]{Chris Lowder}
\affiliation{Southwest Research Institute, Boulder, CO, USA}
\email{}
\author[orcid=0000-0002-6061-6443]{Derek Lamb}
\affiliation{Southwest Research Institute, Boulder, CO, USA}
\email{}

\begin{abstract}
    The transition from sub-Alfv\'enic flow in the magnetically-controlled solar corona to the super-Alfv\'enic solar wind is believed to occur across an extended region of space that has been termed the Alfv\'en zone. In this study we employ new remote sensing observations from NASA's {Polarimeter to Unify the Corona and the Heliosphere} (PUNCH) to probe the upper edge of this region (\(\sim23\,\dash\,32\,\rs\)). We apply Fourier motion filtering to time sequences of white-light images to separate the dominant outbound flow from inbound (sunward) motions, with the latter serving as an indicator of a sub-Alfv\'enic medium. The PUNCH field of view spans all (projected) solar latitudes and enables an assessment of inbound flow outside the ecliptic, thereby overcoming a major limitation of in-situ measurements. The results indicate the occasional presence (filling fraction \(\sim0.19\)) of coherent inbound flows in this region, with characteristic speeds that are consistent with expectations based on modeling and in-situ observations of the Alfv\'en zone. The approach demonstrated in this initial study will be extended in future comprehensive analyses that will aim to map the global structure of the Alfv\'en zone based on refined and expanded PUNCH datasets.
\end{abstract}


\section{Introduction}

The Sun's corona is not gravitationally confined, but expands to form the
supersonic solar wind filling the solar system \citep{Parker_1958}. 
The wind ultimately becomes super-Alfvénic, thereby exceeding the speed of all the 
low-frequency magnetohydrodynamic (MHD) wave modes \citep{weber1967ApJ148}. 
The locus where this happens --- 
in one-dimensional (1D) treatments the ``Alfv\'en point'' \citep{hundhausen1972coronal} 
or in three-dimensional (3D) treatments the ``Alfv\'enic surface'' \citep{pneuman_kopp_1971},
``heliobase'' \citep{zhao_hoeksema_2010}, ``Alfv\'en radius''  \citep[e.g.,][]{goelzer2014JGR},
``Alfv\'en surface'' \citep{deforest2014ApJ787}, or ``Alfv\'en zone'' 
\citep{deforest2018ApJ} ---
is a natural boundary separating the corona from the interplanetary solar wind.  
Inside the Alfv\'en
surface, disturbances in the plasma can, in principle, propagate throughout the entire corona. 
Outside the surface, disturbances in the plasma cannot (at least in the MHD approximation)
propagate sunward. 

The Alfv\'en surface is important for many reasons:  in addition to providing a natural
upper boundary for the corona, it regulates spin-down torque on stars as a whole 
\citep{mestel_1968}, controls open/closed topology in the corona \citep{hundhausen1972coronal},
defines a source surface for the interplanetary magnetic field \citep{goelzer2014JGR}, and 
regulates formation of turbulence in the still-accelerating solar wind \citep{verdini2007apj}.

With the advent of sensitive remote imaging of the outer corona, a natural line of inquiry has
been to search for explicit signs of the Alfv\'en surface's location.  The SOHO/LASCO 
coronagraph \citep{brueckner_etal_1995} suite afforded low-noise observation of the outer
corona.  Using those data, \citet{sheeley_etal_1997} and \citet{wang_etal1999} observed
inbound features below 6~R$_\odot$, concluding that the Alfv\'en point was higher than that.
\citet{verdini2009ApJ} demonstrated, through simulation, the effect of the Alfv\'en surface
on the dynamics of small features (either bulk plasma or wave fronts).  
\citet{deforest2014ApJ787} used STEREO/COR2 data \citep{howard_etal_2008} to search for an 
Alfv\'en surface to the outer limit of COR2; they observed inbound features at altitudes as
high as 13 R$_\odot$, concluding the Alfv\'en surface was at least that high.  Based on the 
shape
of the observed ``speed spectrum'', \citet{tenerani_velli_deforest_2016} concluded that the 
inbound features were most likely plasmoid reconnection exhausts rather than waves.  Either 
type of feature limits the location of the Alfv\'en surface. \cite{deforest2018ApJ} analyzed the density structure of the outer corona and concluded that the Alfv\'en (and 
therefore fast-mode) speed is so varied that there is likely no single identifiable smooth
critical surface, but rather an ``Alfv\'en zone'' of gradual transition from coronal to 
solar-wind-like MHD characteristic vectors.  This is corroborated by modeling work by 
\citet{Chhiber2022MNRAS} which showed the development of super-Alfv\'enic and sub-Alfv\'enic 
``islands'' in the flow through the outer corona.  

Parker Solar Probe \citep[PSP;][]{fox2016SSR,Raouafi2023SSR}
is specifically designed to probe the corona itself in situ, and indeed it passed
through the Alfv\'en zone/surface and verified its presence with a suite of in-situ 
measurements of the plasma condition \citep{kasper2021prl}; analysis of the 
effect of the Alfv\'en zone on the plasma condition, using PSP data, is an 
ongoing effort \citep[e.g.,][]{bandyopadhyay2022ApJ,Chhiber2024MNRAS,Adhikari2026ApJ}, and at 
least some of the expected ``islands'' have been seen, associated with coronal
mass ejections \citep[CMEs;][]{jagarlamudi_etal2025}, in addition to deformed-surface effects from 
plasma conditions in the corona \citep{ervin_etal024,Badman2025ApJ}. 

One science objective of the Polarimeter to Unify the Corona and Heliosphere (PUNCH) mission
\citep{DeForest2026SoPh} is to characterize the shape and evolution of the Alfv\'en zone
via tracking of inbound features in the critical region between 7 and 50 R$_\odot$ from the 
Sun itself, using image-processing methods to isolate inbound features.  An overview is 
available
from \citet{Cranmer2023SoPh}. Importantly, PUNCH's circular field of view (FoV) spans all heliolatitudes (projected on to the image plane) and permits characterization of the Alfv\'en zone outside the ecliptic plane, thereby overcoming a major limitation of in-situ spacecraft datasets \citep[cf.][]{Verscharen2021MNRAS}.

In this work, we report on an initial analysis of data from PUNCH, 
to replicate and extend the results of \cite{deforest2014ApJ787}, shifting our attention to the outer edge of the Alfv\'en zone (\(\sim 23\,\dash\,32\,\rs\)) where inbound flow may be less widespread.
In the following sections, we describe the relevant PUNCH data products and a reduction 
method to extract inbound features from the PUNCH mosaic data; present results from 
carrying out that analysis on PUNCH preliminary (version 0l) data; and discuss and draw conclusions about the nature of the Alfv\'en zone and its measurability with PUNCH -- both 
with the current preliminary data and with the anticipated version-1 reduced data set.

\section{Data and Methods}\label{sec:data}
\subsection{Data Selection, Coordinates, and Pre-processing}
We used version v0l of the PUNCH Level 3 Clear-Averaged-Mosaic (L3 CAM) data product, whose preparation includes background (F-corona) subtraction and starfield suppression. These images are 32-min averages of total brightness that fill the entire circular FoV. The data were downloaded in FITS format from NASA's Solar Data Analysis Center \citep[SDAC;][]{Ireland2025SDAC},\footnote{ \url{https://umbra.nascom.nasa.gov/punch/3/CAM/}} and cover the time period from 2026 May 01 00:16:00 to 2026 May 06 23:44:00 (UTC). The IDL program {\fontfamily{cmtt}\selectfont read\_punch.pro}, made available by the PUNCH mission SOC,\footnote{PUNCH data analysis software is available at the mission's \href{https://github.com/punch-mission/PUNCH_IDL_Tools}{github}.} was employed to read the .FITS files and obtain image headers and data. 

\begin{figure}
    \centering
       \includegraphics[width=.4\textwidth]{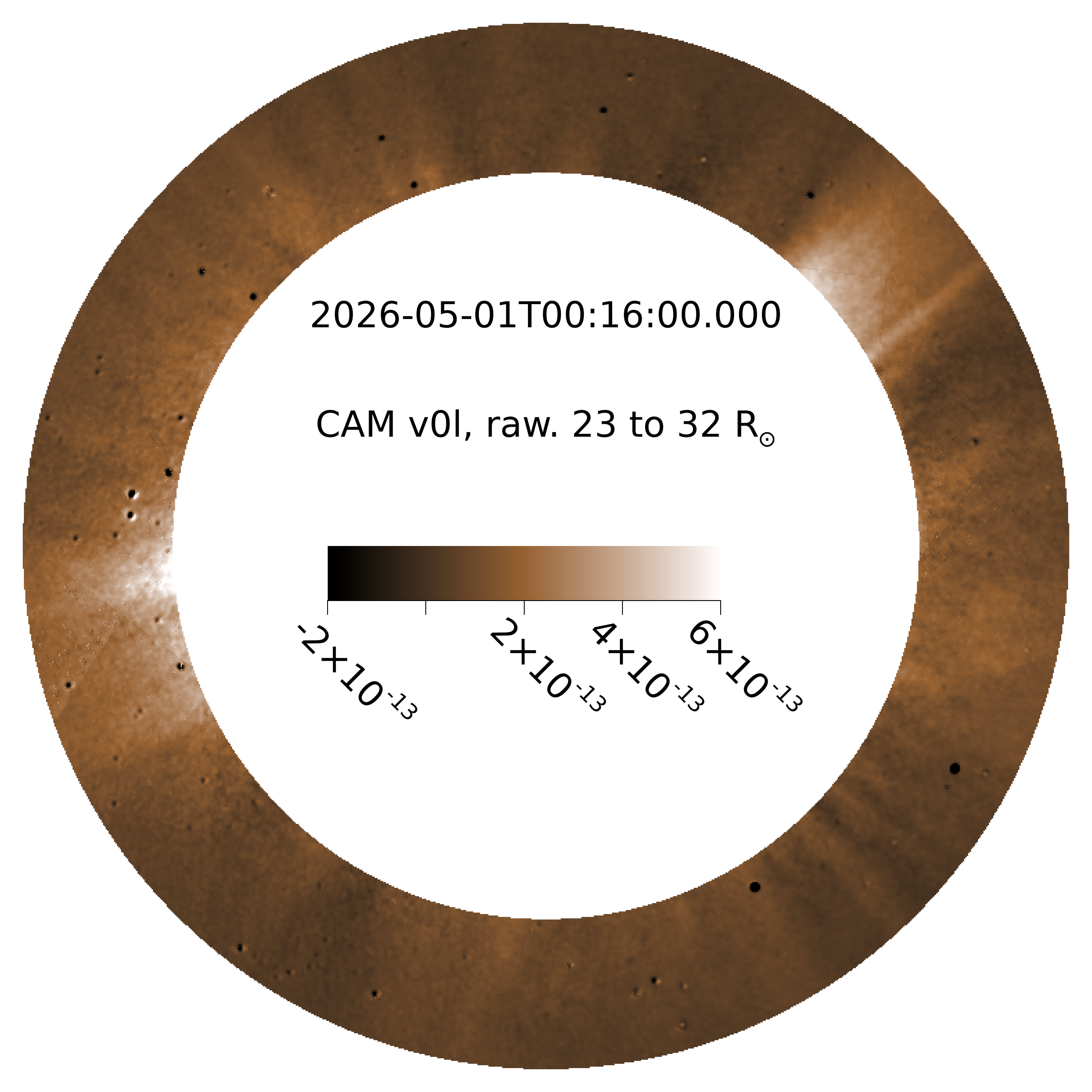}
       \includegraphics[width=.4\textwidth]{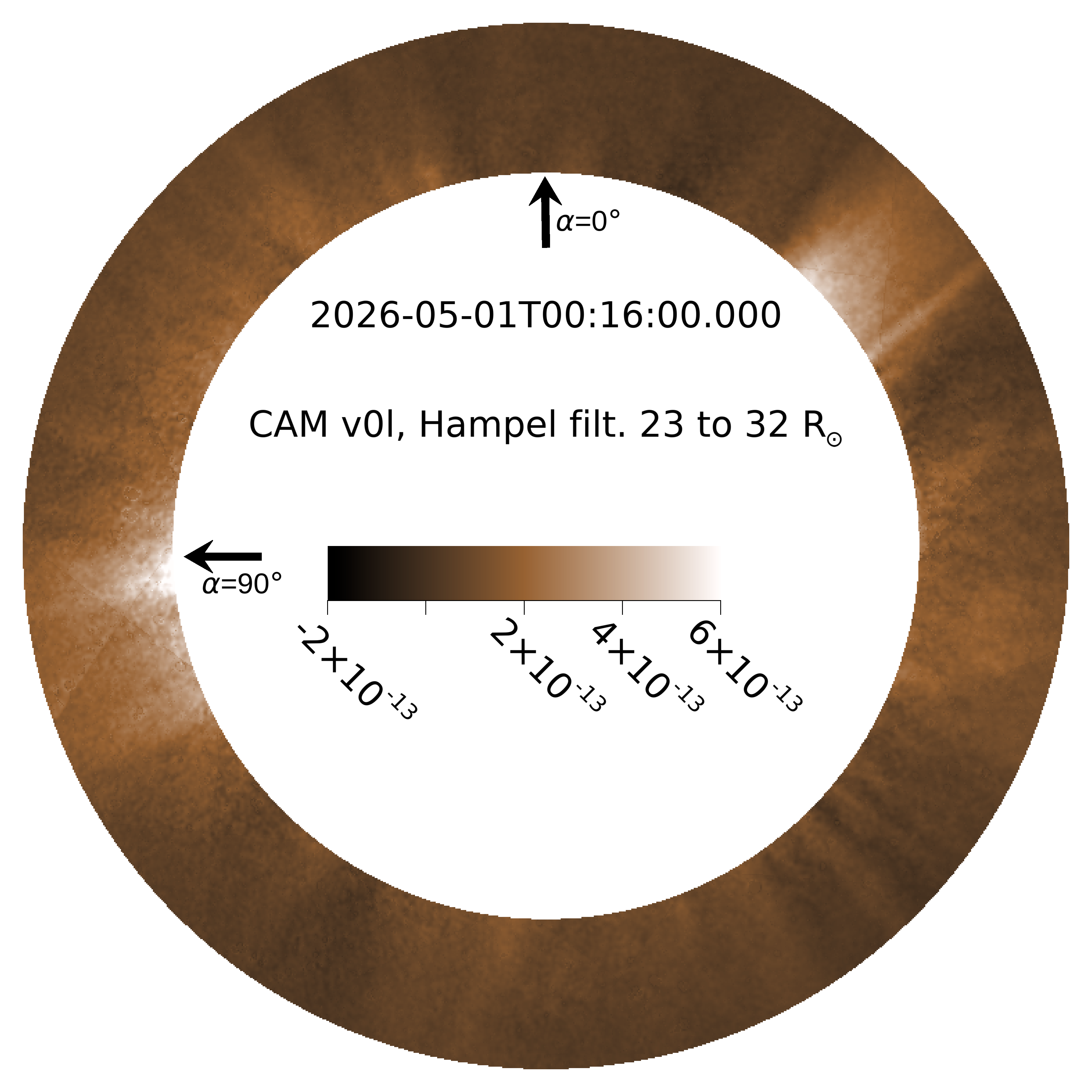}
    \caption{\textit{Top panel} shows a sample PUNCH CAM v0l image of total brightness, cropped to the region of interest (\(23\,\dash\,32\,\rs\)). Colorbar indicates brightness in units of MSB. \textit{Bottom panel} is produced by application of the Hampel filter to remove point-like artifacts (see text and App. \ref{sec:Hampel}). Arrows indicate the azimuthal angle (\(\alpha\)) convention: increasing CCW starting from solar north. Movie files are included for both panels in Supplementary Material.}
    \label{fig:Hampel}
\end{figure}

SolarSoft WCS routines \citep{Freeland1998SoPh} provided helioprojective cartesian (gnomonic projection, or {\fontfamily{cmtt}\selectfont TAN}) coordinates of each image pixel: \(\theta_x,~\theta_y\), in degrees \citep{Thompson2006AA,DeForest2026SoPh}. From these we estimated the heliocentric radial coordinate:
\begin{equation}
    r = \sqrt{\theta_x^2 + \theta_y^2} \times 3600 / R_{\odot,\text{arcsec}},
\end{equation}
where multiplication by 3600 converts degree to arcsecond, and \(R_{\odot,\text{arcsec}}\) is a parameter obtained from the FITS header that expresses a solar radius in arcseconds.\footnote{Here we have used the small-angle approximation for elongation \(\varepsilon=\sqrt{\theta_x^2+\theta_y^2}\) \citep{deforest_etal2026} since we restrict our analysis to \(6^\circ<\varepsilon<8.5^\circ\), corresponding approximately to \(23~\rs<r<32~\rs\).} The azimuthal angle is defined (in degrees):
\begin{equation}
    \alpha=\arctan(-\theta_x/\theta_y) \times \frac{180\degree}{\pi},
\end{equation}
which is wrapped modulo 360\degree~onto the interval \([0\degree,360\degree)\). This angle is zero at solar north (``up'' direction in Fig. \ref{fig:Hampel} images) and increases counter-clockwise (CCW) so that 90\degree~aligns with solar east (``left'' in Fig. \ref{fig:Hampel}).

PUNCH science data are currently (as of 09/2026) preliminary (v0xx) and the v0l data only include the WFI instruments \citep{Laurent2026SoPh}, which at present cover \(r \gtrsim  18~R_\odot\). Image data between \(18\, \dash \, 22\, \rs\) are significantly affected by processing artifacts, so we focus on the radial band between 23 and \(32\,\rs\) (corresponding to elongations of \(\sim 6^\circ\) and 8.5\degree), which is expected to overlap with the upper edge of the Alfv\'en zone \citep[based on prior modeling and in-situ observations;][]{Chhiber2022MNRAS,Chhiber2024MNRAS,Badman2025ApJ}. Data outside these bounds are discarded. An example of the resulting images is shown in the top left panel of Fig. \ref{fig:Hampel}. The time period analyzed is from an epoch of high solar activity \citep{Usoskin2023LRSP_sol_activity}, and the FoV samples a couple of eruptive events (see movies in Supplementary Material).

Although we have used the current data version, the preliminary science images contain artifacts related to data processing steps such as starfield suppression and merging of images from the three WFI cameras to form the CAM mosaic \citep{DeForest2026SoPh}. To remove point-like artifacts that can be classified as statistical outliers we apply a two-dimensional (2D) generalization of a Hampel filter \citep{Pearson2016EJASP}, which classifies individual pixels as outliers based on their median deviation from the local median brightness; such outliers are replaced with the local median value. See App. \ref{sec:Hampel} for details. Fig. \ref{fig:Hampel} shows an example of this filtering, which is evidently quite effective in removing point-like artifacts without affecting ``good'' pixels (see also Fig. \ref{fig:Hampel2}).

The temporal cadence of the images is 32 minutes, but there are a handful of isolated (never consecutive) missing frames. To obtain a sequence of images with a constant temporal cadence (for subsequent Fourier transformation) we employ linear interpolation across missing frames. Additionally, we isolated (by eye) three frames in our image sequence wherein large artifacts pollute a major fraction of the image area; such frames, which are also isolated in occurrence, are treated as missing and similarly interpolated across.

The image data are resampled to a uniform grid in \(r\) and \(\alpha\) 
using natural-neighbor interpolation \citep{watson2013contouring}. The uniform grid has \(\Delta r = 0.035\,\rs\) and \(\Delta\alpha=0.70\degree\), producing a resampled image of size \(512\times256\) in \(\alpha\)-\(r\). This resampling is done for each image corresponding to the 270 timestamps. 
\subsection{Radial Filtering and Fourier Analysis}
\begin{figure}
    \centering
    \includegraphics[width=.5\textwidth]{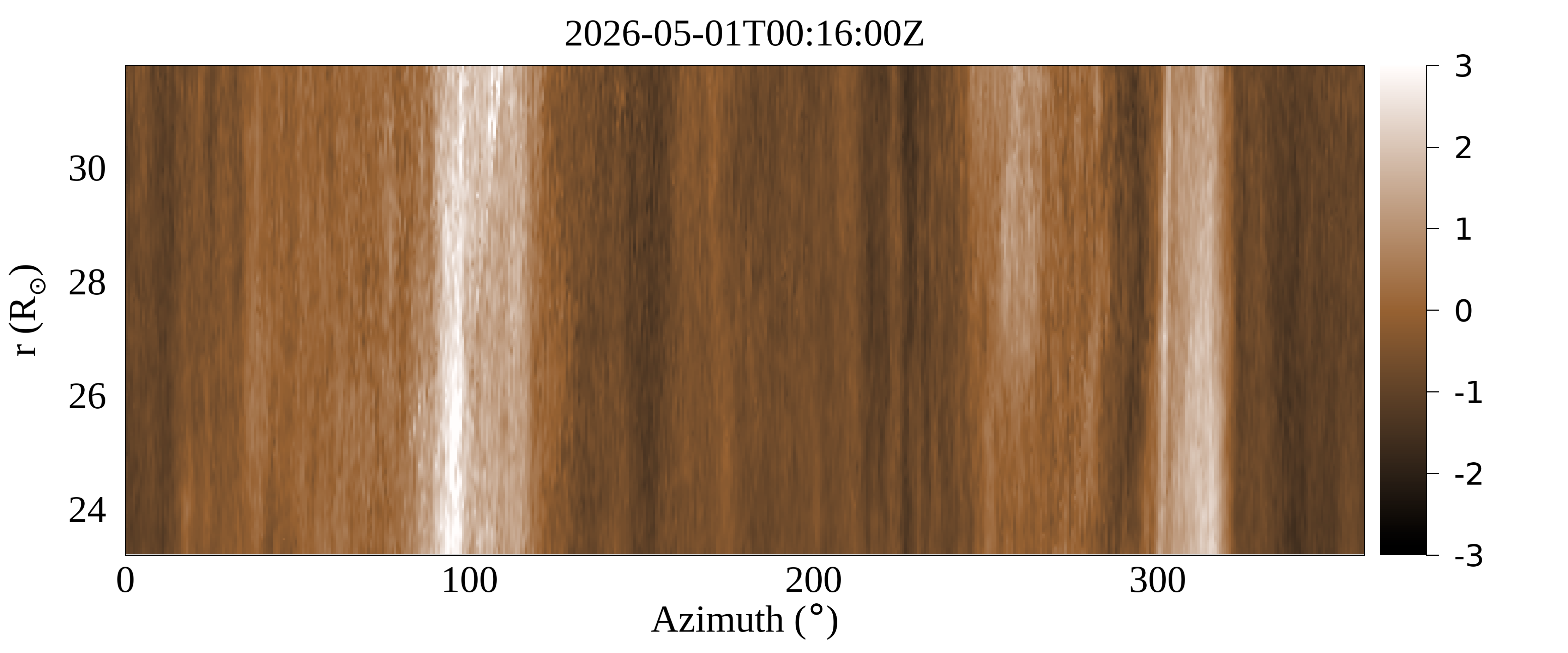}
    \caption{Example radially-filtered image, corresponding to the bottom panel of Fig. \ref{fig:Hampel}, produced as described in the text. A movie file is included in Supplementary Material.}
    \label{fig:radfilt}
\end{figure}
The time sequence of radius vs azimuth images forms a 3D data cube. A radial filter is applied to the images by (i) subtracting a data cube containing the mean of the original data across azimuth and time, and (ii) dividing by a data cube containing the standard deviation computed across azimuth and time. 
This removes the ambient radial and temporal gradients in radiance, giving access to radial motion of differential signals \citep{deforest2014ApJ787}. An example image following such radial filtering is shown in Fig. \ref{fig:radfilt}, which corresponds to the image shown in Fig. \ref{fig:Hampel}. 

We then apply a motion filter based on Fourier transformation of the image sequence, to separate inbound (sunward) and outbound (anti-sunward) features  \citep{deforest2014ApJ787,Cranmer2023SoPh}. The images are first apodized in the radial and temporal domains with a cosine taper applied to the first and last 10\% of each domain. Then a forward FFT is performed in the \(r\) and temporal dimensions, yielding, for each \(\alpha\), a \(k_r\dash\omega\) spectrum. Moving features are localized by speed and direction in the \(k_r\dash\omega\) plane, with speed \(V=-\omega/k_r\). Regions in the \(k_r\dash\omega\) plane where \(k_r \omega >0\) and \(k_r\omega<0\) correspond to inbound and outbound motions, respectively. To separate these motions, we create `in' and `out' masks that ``zero out'' the outbound and inbound quadrants in the \(k_r\dash\omega\) plane, respectively. Following \cite{deforest2014ApJ787}, we also apply  masks that zero out motions with \(900\,\kmps < V < 19\,\kmps\), as well as the \(k_r=0\) line where \(V\) diverges. This further suppresses, to a large extent, near-stationary (slowly moving) structures like stars. The Fourier filter also reduces image flicker that is present in the preliminary PUNCH data; this flicker comes from rapid temporal fluctuations in the overall brightness of an image, and is therefore associated with low \(k_r\) (since it affects an image globally) and high \(\omega\) (since it occurs at the 32-min temporal resolution of the data). For further details on the theory of Fourier motion filtering in the context of heliophysics imaging, see \cite{deforest2014ApJ787} and references within.

The compounded Fourier-domain masks are applied, producing two datacubes -- outbound and inbound -- and reverse FFTs are performed on each to obtain the respective time sequences of \(\alpha\dash r\) images. Note that noise sources in the resulting images can be of an isotropic type (e.g., photon noise), or those related to the Fourier-domain speed cutoffs. The former type of noise would be equally present in both inbound and outbound portions of the data, and can be eliminated by searching for coherent (non-random) structure in the filtered data. Artifacts related to the Fourier cutoffs can be avoided by focusing on a physically-sensible range of speeds that are well separated from the characteristic filter speeds (see Sec. \ref{sec:Result}, below). 

\begin{figure}
    \centering
      \includegraphics[width=.5\textwidth]{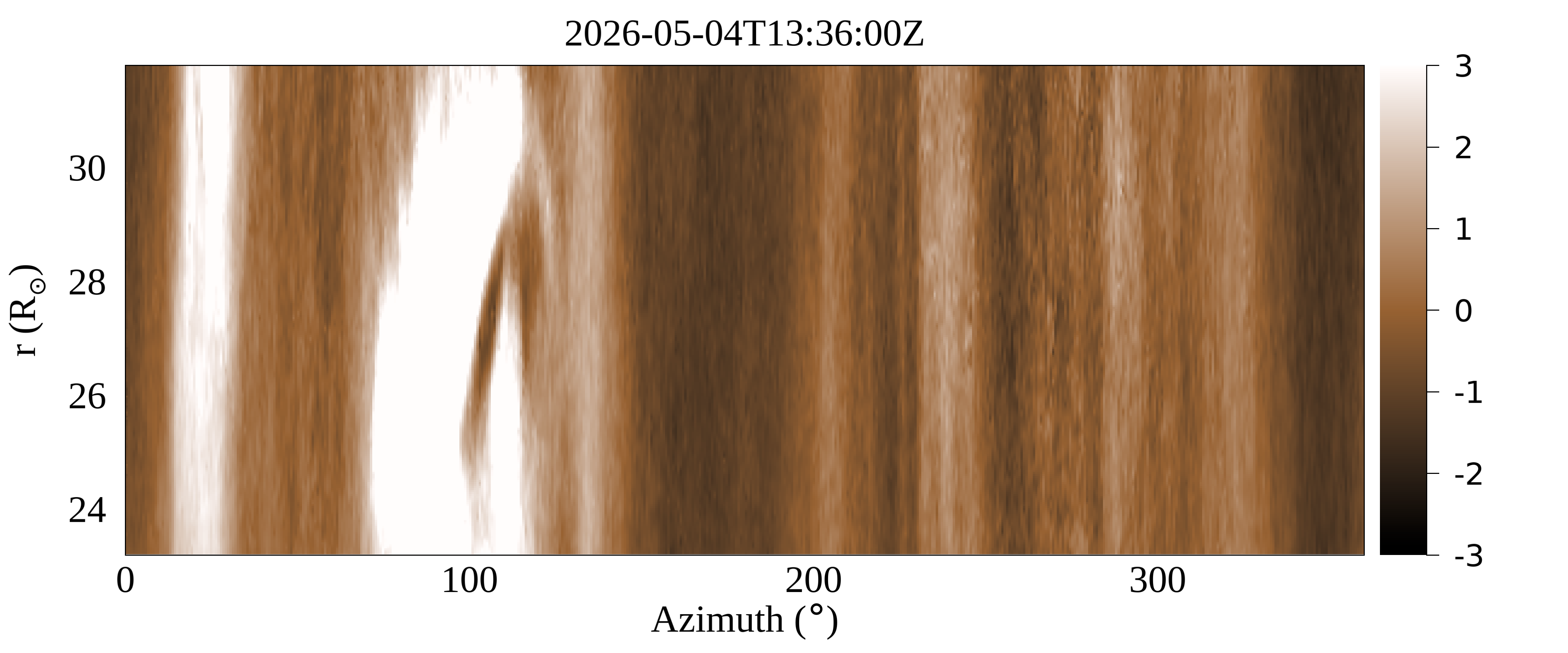}
       \includegraphics[width=.5\textwidth]{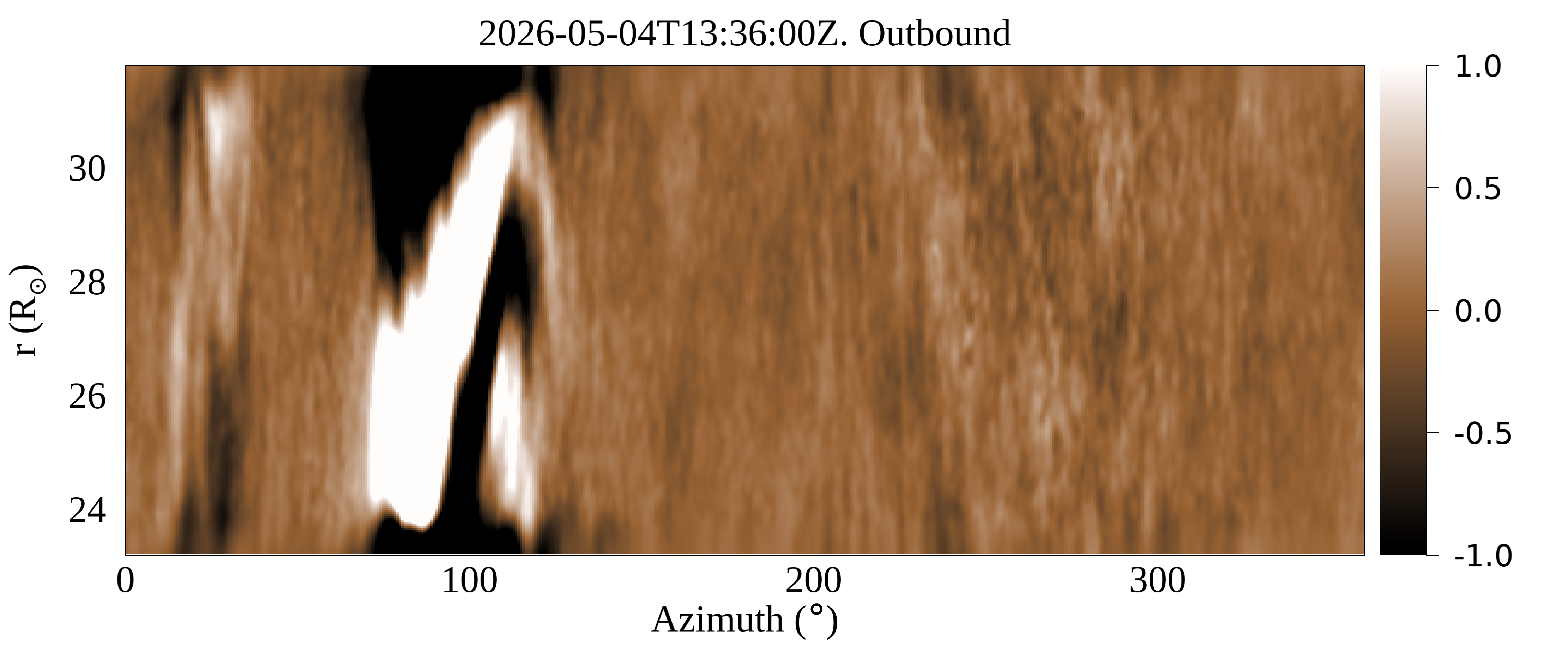}
       \includegraphics[width=.5\textwidth]{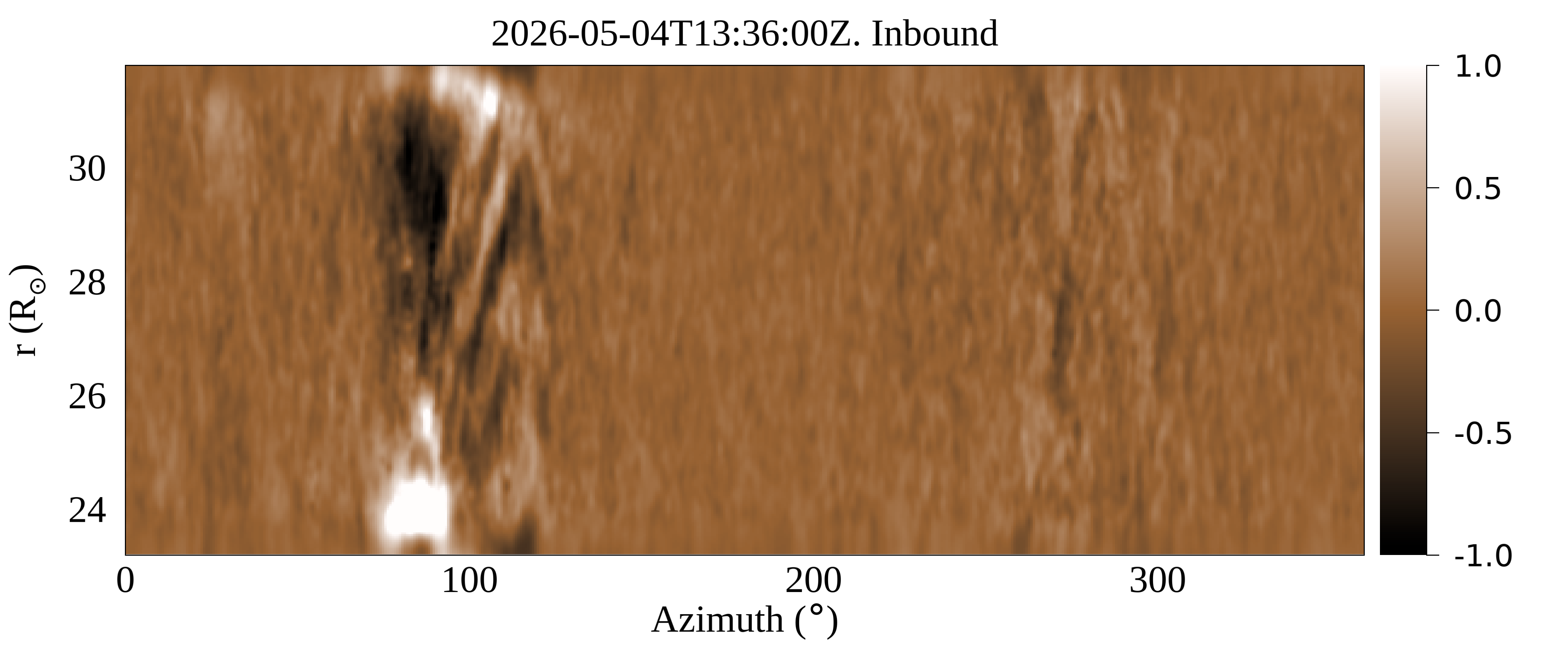}
    \caption{Middle and bottom panels show outbound and inbound features corresponding to the example image in the top panel, following Fourier motion filtering as described in text. Movie files are included for all panels in Supplementary Material.}
    \label{fig:inout}
\end{figure}

Fig.~\ref{fig:inout} shows example outbound and inbound \(r\dash\alpha\) images produced in this way. They correspond to the (pre-Fourier-filtering) image shown in the top panel. Gaussian smoothing has been applied to the filtered images, 
using a window of size 2.5\degree~in \(\alpha\) and \(0.2\,\rs\) in \(r\). Movie files are available in the Supplementary Material. The outbound image shows the reduction in quasi-stationary structure compared to the top panel (e.g., in the feature around \(\sim 20\)\degree). A CME is in progress at around 100\degree. The Fourier-filtered outbound movie shows a reduced level of flicker compared to the pre-filtering case, but the movies still indicate some flicker-type noise around azimuths of 90\degree~and 270\degree, i.e., at low heliolatitudes; this should be kept in mind while interpreting results from these regions.

To further evaluate inbound flow characteristics, we proceed to the main Results section below, wherein we separately examine flows associated with individual azimuths via analysis of \(r\dash t\) (or \textit{height-time}) plots.
\section{Results}\label{sec:Result}
\subsection{Height-Time Images}\label{sec:r-t}
\begin{figure*}
    \centering
      \includegraphics[width=.65\textwidth]{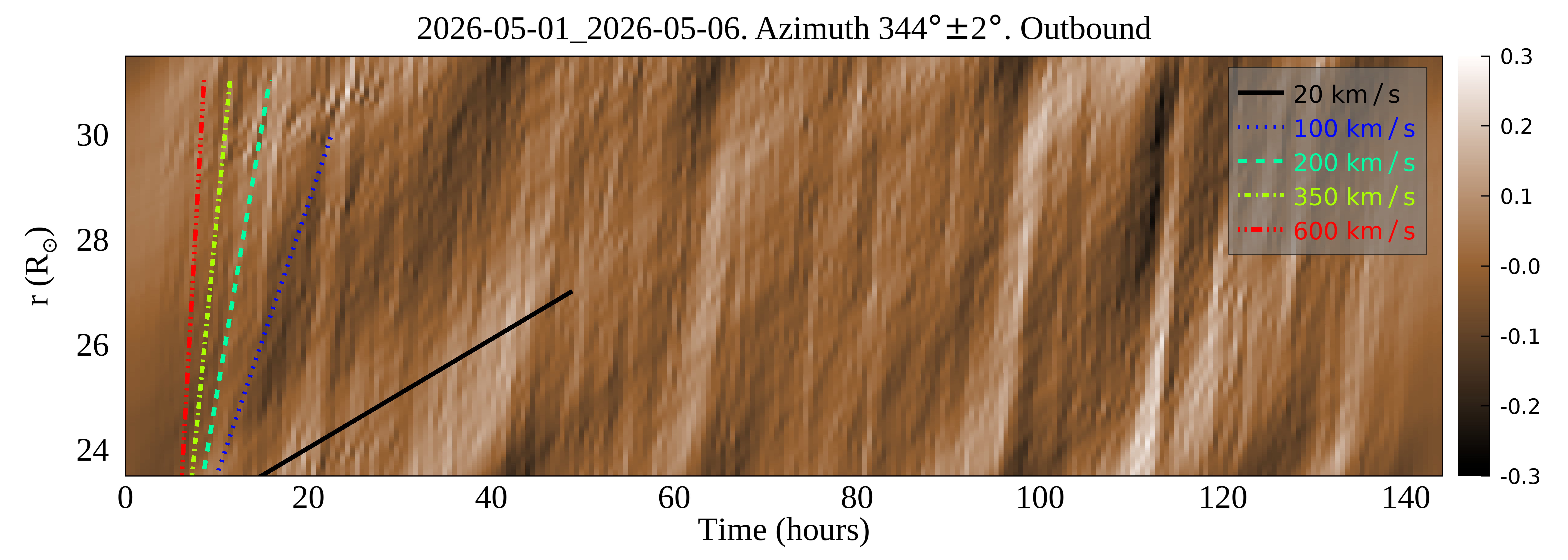}
      \includegraphics[width=.65\textwidth]{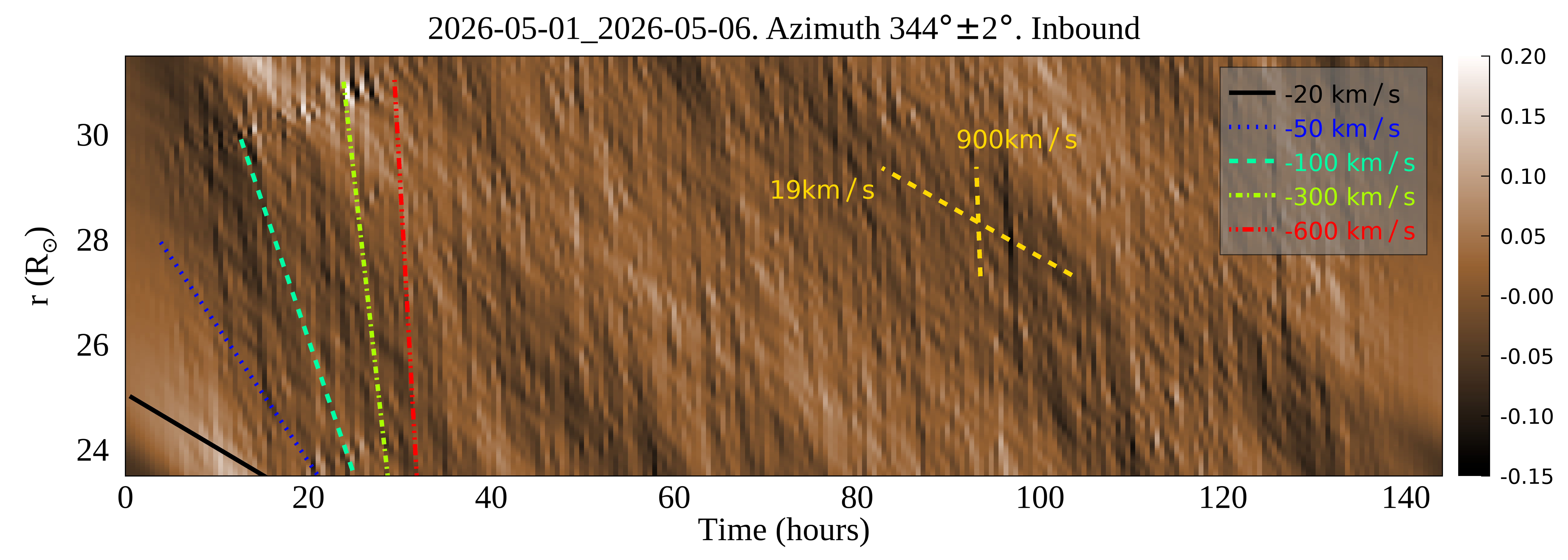}
      \includegraphics[width=.65\textwidth]{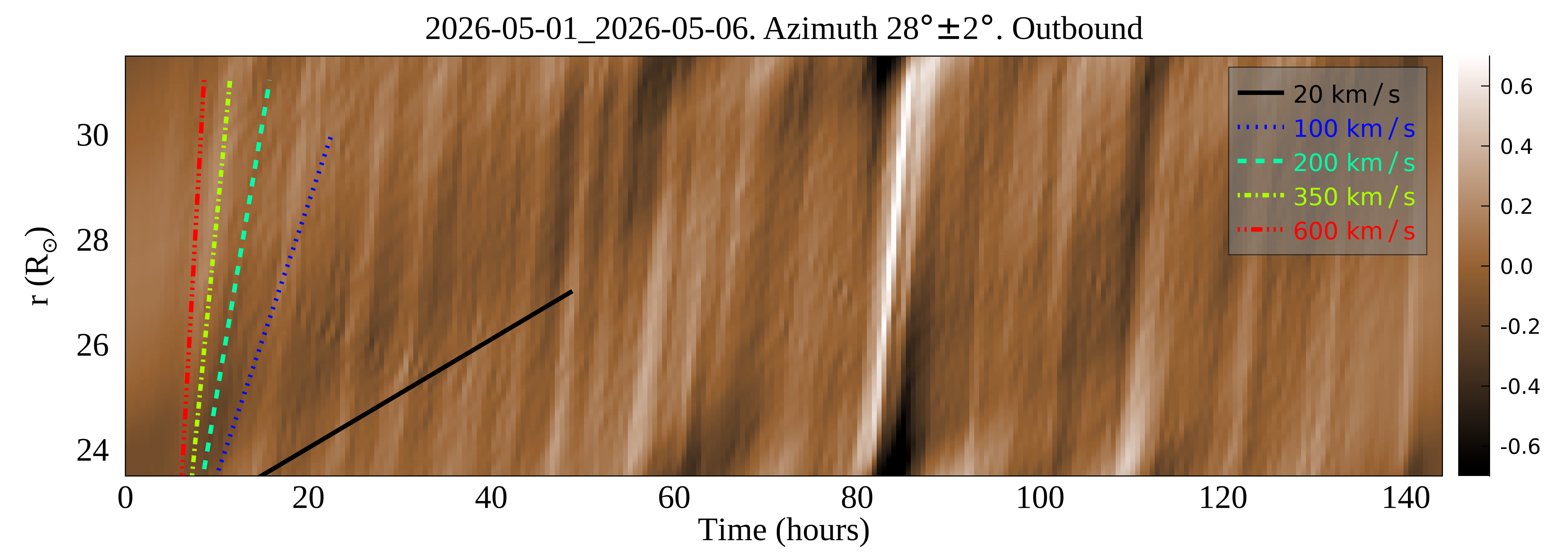}
      \includegraphics[width=.65\textwidth]{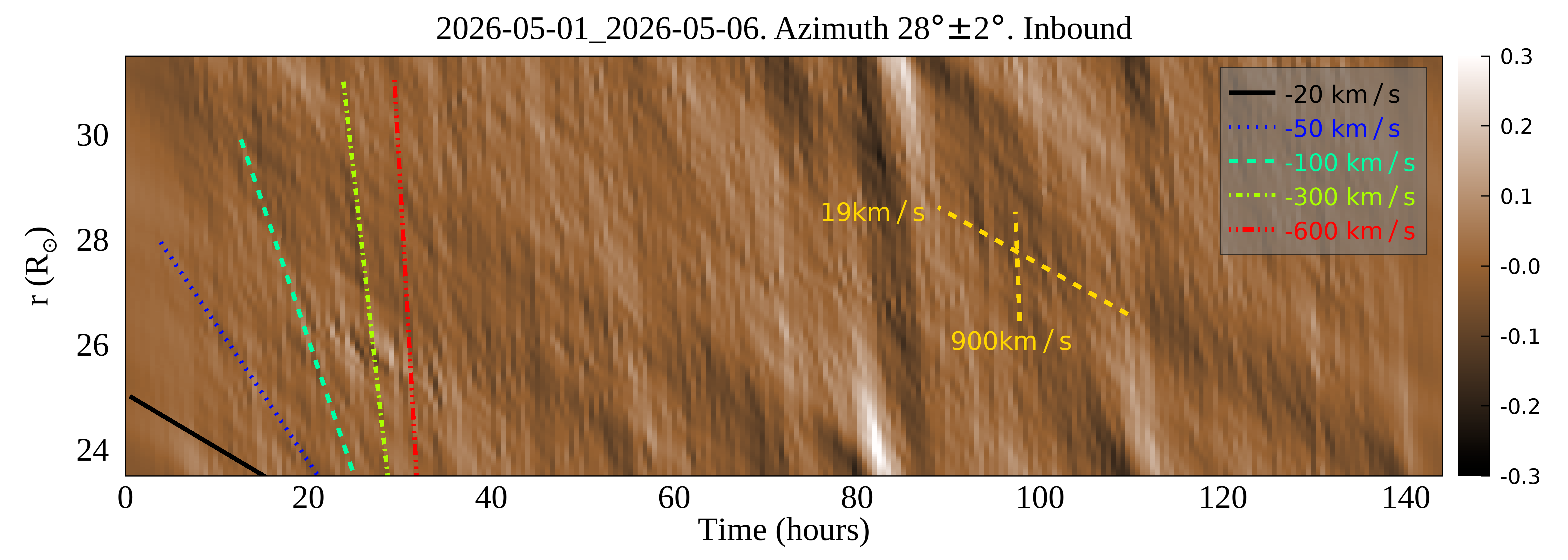}
    \caption{Radius vs time plots at fixed azimuths (\textit{top two panels}: 344\degree; \textit{bottom two panels}: 28\degree; each averaged over \(\pm2^\circ\)) for outbound and inbound cases. Reference slopes corresponding to listed speeds are overlaid. Dashed gold lines in inbound panels indicate Fourier speed-filter cutoffs at \(19\,\kmps\) and \(900\,\kmps\).}
    \label{fig:r-t}
\end{figure*}

The temporal and radial evolution of two selected azimuthal slices (at 28\degree~and 344\degree) is shown in Fig. \ref{fig:r-t}, separately for the outbound and inbound cases; here the slices are averaged over 2\degree~of azimuth. The speed of a moving feature can be estimated from the associated slopes in these \(r\dash t\) images. Reference slopes are overlaid, with corresponding speeds annotated in legends. Recall that the Fourier speed-filter cutoffs are at \(19\,\kmps\) and \(900\,\kmps\). 

The outbound cases indicate clear evidence of coherent flows, whose speeds can be roughly estimated from their slopes to be around 200 to 400 \(\kmps\); values that are well-isolated from the speed cutoffs. There is also a suggestion of acceleration, indicated via the increasing slope of some ridges, e.g., just after 60 hours in the 344\degree~azimuth case. This will also be apparent in the associated speed spectrum (Sec. \ref{sec:spect}, below).

The \(r\dash t\) images for the inbound case are similar to those obtained by \cite{deforest2014ApJ787} from an analysis of STEREO/COR2 data between \(\sim 3~\dash~ 15~\rs\). Since the range of distances covered in the present work is expected to be near the upper edge of the Alfv\'en zone, we may anticipate that coherent inbound flows will not be ubiquitous; we will attempt a systematic evaluation of their filling fraction in Sec. \ref{sec:spect}. A visual inspection of the images indicates myriad diffuse downward ridges of varying slope and durations between \(\sim 2\) and \(\sim 10\) hours. While several features have slopes that are distinct from the filter cutoffs and appear to be sufficiently coherent (i.e., long-lived) to be distinguishable from noise, in the next section we conduct an image-convolution analysis of the \(r\dash t\) plots to better identify coherent signals and to estimate prevailing flow speeds.
\subsection{Speed Spectra}\label{sec:spect}
For illustration, we first describe the speed spectrum calculation for both outbound and inbound flow images at 344\degree~azimuth. The approach is similar to that described in \cite{deforest2014ApJ787}. We begin with the 344\degree~image since it is relatively isolated from large transient flows (likely associated with CMEs) and better associated with ambient solar wind. For the outbound case, we isolate an image patch spanning \(4\,\rs\) and 50 hours, and convolve such a patch with a straight-line kernel with slope equivalent to a particular speed between \(\sim25\) and \(700\,\kmps\).\footnote{The kernel dimensions are \(\sim 5\) hr and \(\sim 3.5\,\rs\) in the time and radius domains, respectively. The convolution response is set to zero where the kernel extends beyond the edge of the image patch.} We calculate the rms value of the convolved image, which represents a single value in a 2D distribution of optical flow in speed-altitude space; here the altitude is associated with the central helioradius of the patch. We iterate over both speed and central helioradius of the extracted patch. After generating all rms values, we divide each row (i.e., constant-radius locus) in the distribution by its maximum value, thus normalizing the distribution to a maximum of unity. The process is repeated for five 50-hr intervals with starting times at 10, 30, 50, 70, and 90 hours, and we collate the results into a single distribution by selecting the median value in each speed-radius bin across these five intervals.
\begin{figure*}[]
    \centering
      \includegraphics[width=.32\textwidth]{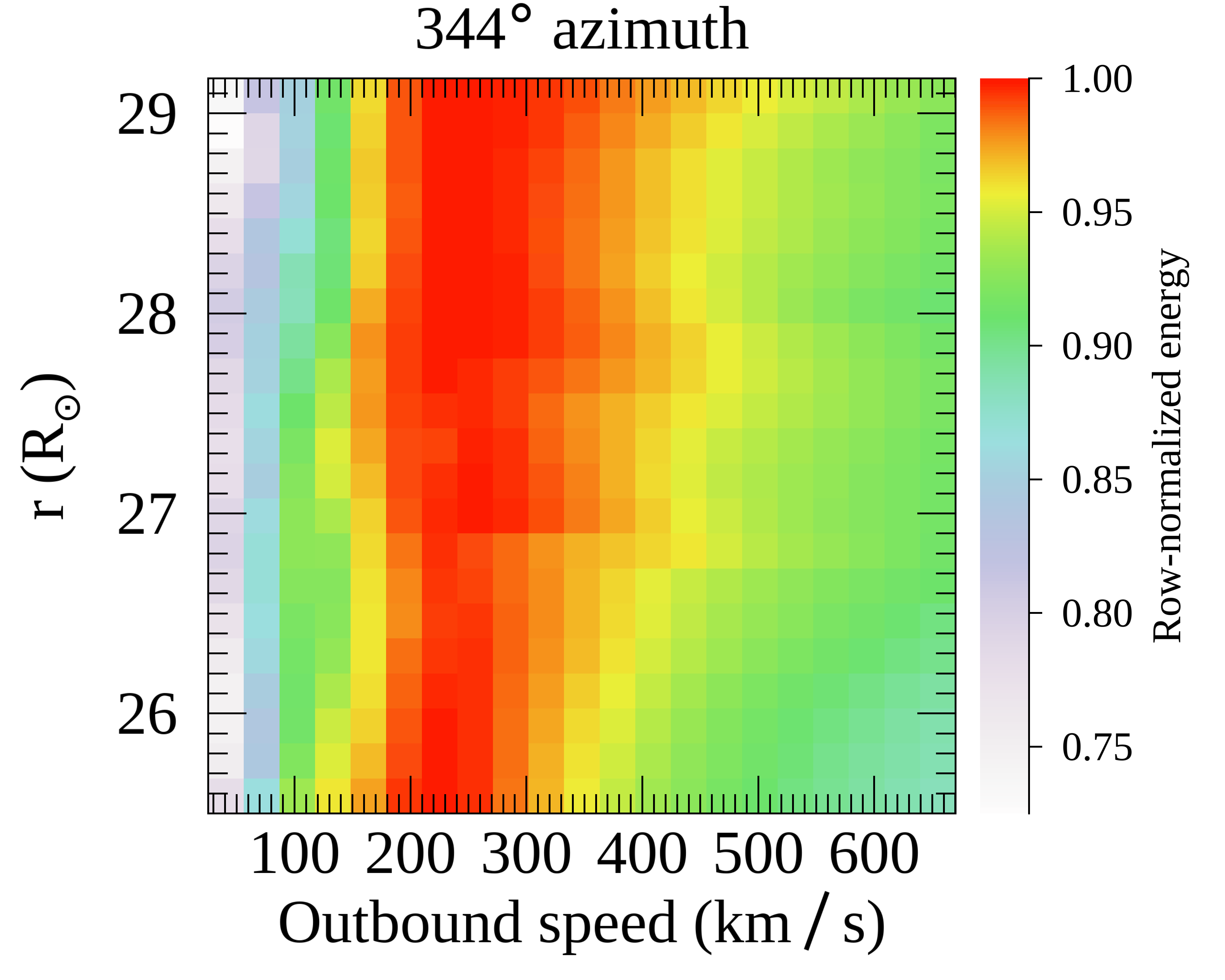}
      \includegraphics[width=.32\textwidth]{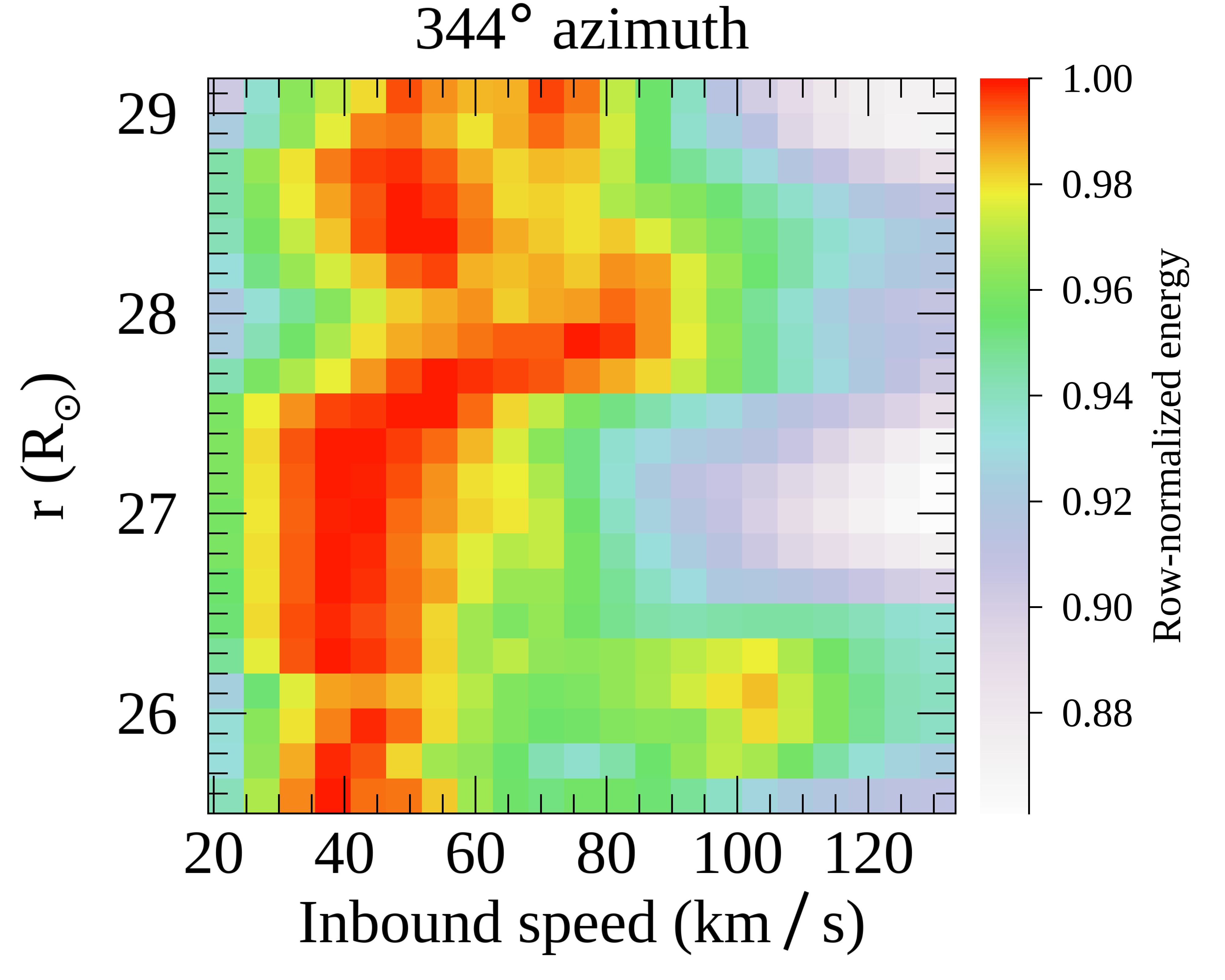}
      \includegraphics[width=.32\textwidth]{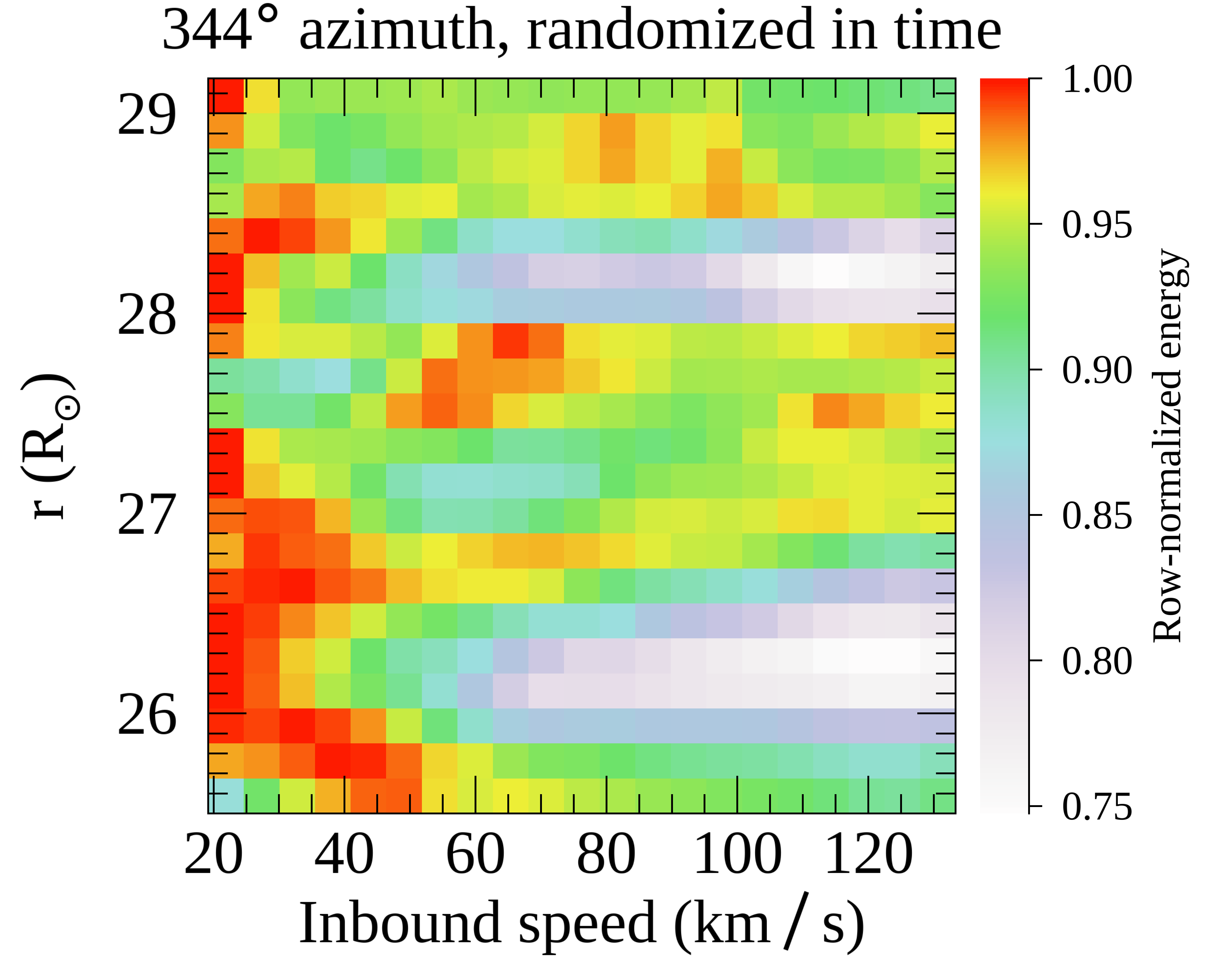}
    \caption{Left and middle panels show speed spectra as a function of altitude, computed as described in the text for the outbound and inbound height-time images at 334\degree~azimuth (top two panels of Fig. \ref{fig:r-t}). Right panel shows the inbound speed spectrum of the corresponding time-randomized image, which is devoid of any coherent flows in time (see text).}
    \label{fig:speed_spect}
\end{figure*}

The resulting speed spectrum in the left panel of Fig.~\ref{fig:speed_spect} indicates a wide ridge where optical flow energy \citep{deforest2014ApJ787} is the highest, centered at speeds of \(\sim200\,\dash\,300\,\kmps\); there is a suggestion of acceleration with \(r\). Such speeds are generally consistent with those estimated from visual inspection of bright ridges in the outbound \(r\dash t\) image, and with expectations from modeling and observations of the young solar wind \citep[e.g.,][]{Liu2024ApJ,Usmanov2025ApJ}.

An analogous convolution analysis of the inbound height-time image produces the speed spectrum shown in the middle panel of Fig. \ref{fig:speed_spect}. The spectrum indicates a wide ridge where optical flow energy is the highest, centered at speeds of \(\sim 40\,\kmps\). No obvious trend with altitude is observed. We estimate expected inflow speeds near the outer edge of the Alfv\'en zone by assuming a typical solar wind flow speed \(V_\text{sw}\sim 300\,\kmps\) at these altitudes \citep[e.g.,][]{Liu2024ApJ}, along with an Alfv\'en Mach number \(M_\text{A} \sim 0.85\) \citep{Chhiber2022MNRAS,Chhiber2024MNRAS,Badman2025ApJ}. We estimate the Alfv\'en speed as \(V_\text{A}\sim V_\text{sw}/M_\text{A}\sim 350\,\kmps\), which implies a typical inbound flow of \(V_\text{A}-V_\text{sw}\sim 50\,\kmps\) in the sub-Alfv\'enic regime. Such values are consistent with the observed optical flow in the middle panel of Fig. \ref{fig:speed_spect}, and the ``broadband'' aspect of the spectrum is consistent with the highly-variable spatial and temporal distribution of \(M_\text{A}\) expected from a turbulent and fragmented Alfv\'en zone \citep{deforest2018ApJ,Cranmer2023SoPh,Chhiber2024MNRAS,Badman2025ApJ}.

To further distinguish the inferred inbound flow from noise, we generate a ``noisy'' image by randomizing the order of the time values in the inbound \(r\dash t\) image at 344\degree~azimuth (Fig.~\ref{fig:r-t}); i.e., the ``columns'' of the image are shuffled randomly. The speed spectrum computed from the resulting image is shown in the right panel of Fig.~\ref{fig:speed_spect}. It indicates a concentration of optical flow at the lowest evaluated speeds (with some scatter), which is expected from an image lacking any flows that are coherent and sustained in time. This provides further evidence that the inferred flows in the inbound image are distinct from random noise. Note that the convolution approach employed here will always yield a finite rms value for any speed, since any straight-line kernel (regardless of slope) will overlap at-least partially with image features \citep{GonzalezWoods2018}.

\begin{figure*}
    \centering
      \includegraphics[width=.246\textwidth]{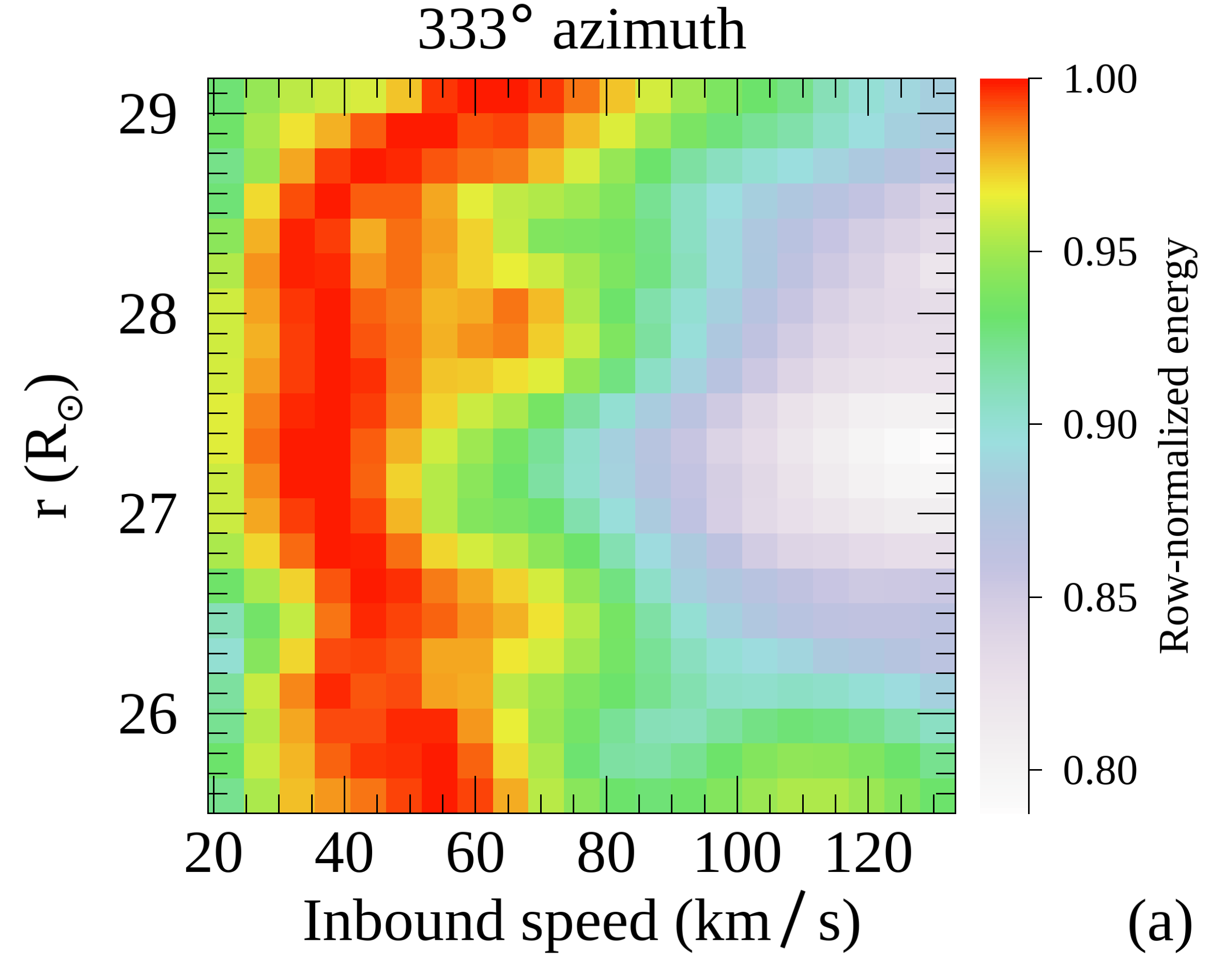}
      \includegraphics[width=.246\textwidth]{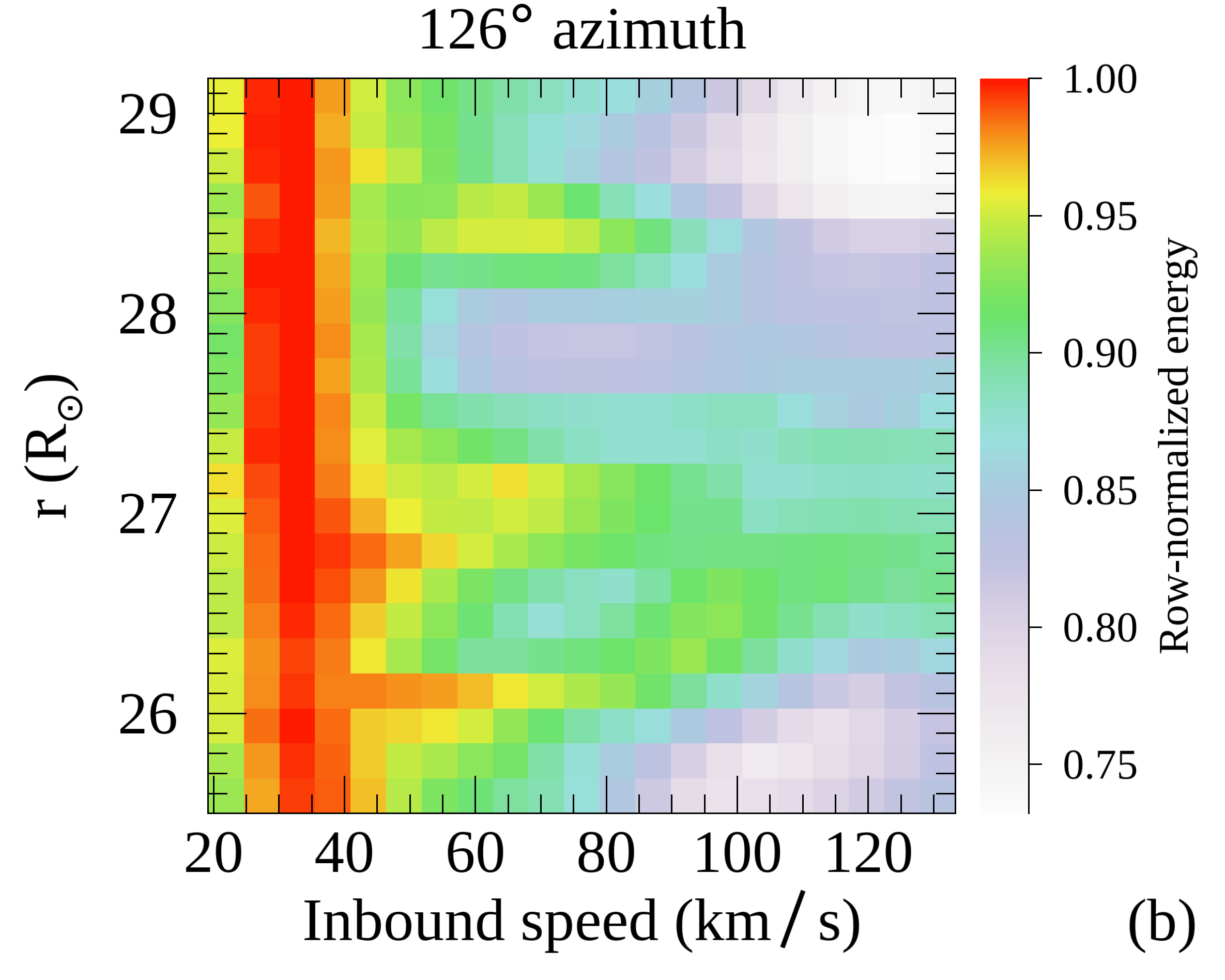}
      \includegraphics[width=.246\textwidth]{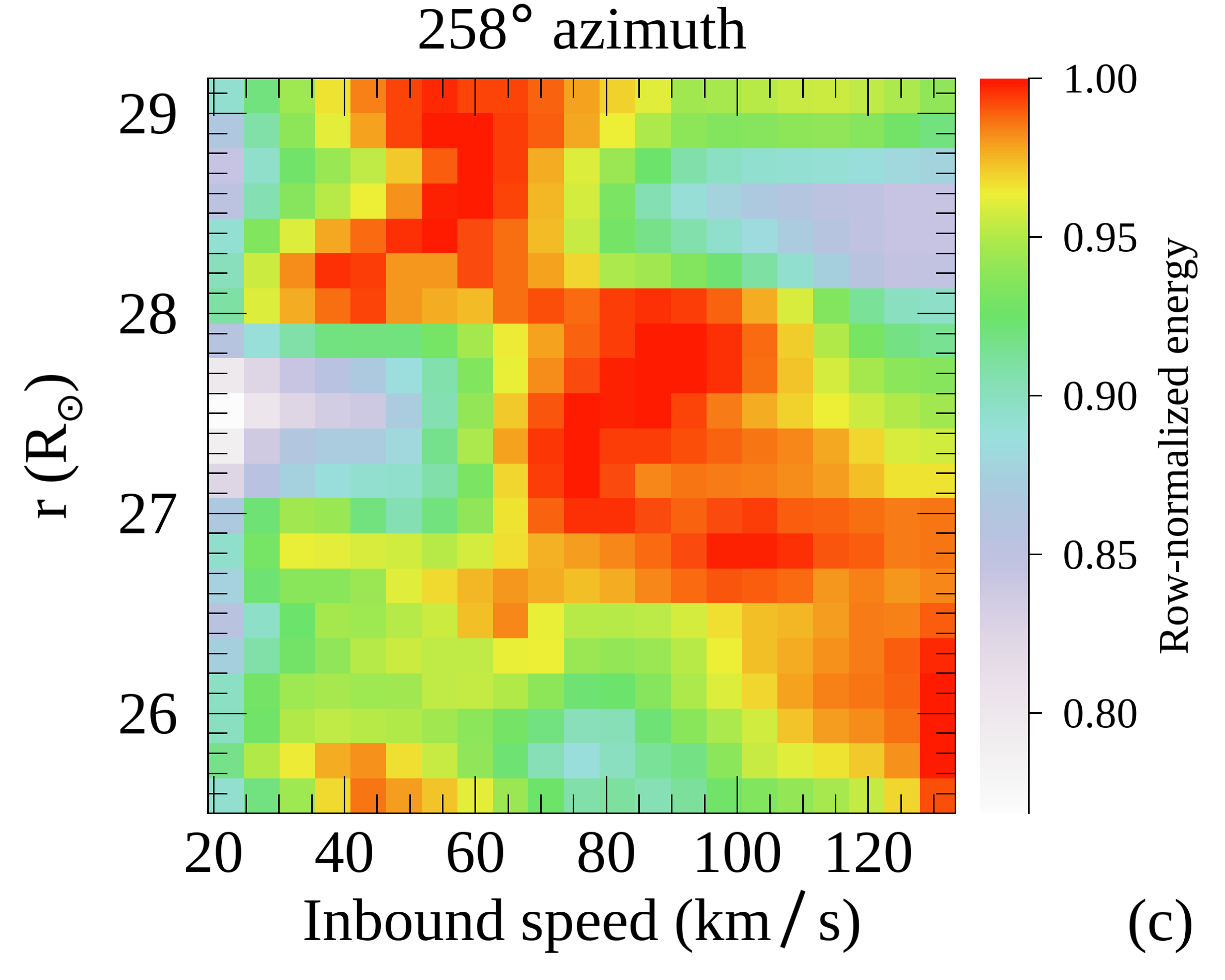}
      \includegraphics[width=.246\textwidth]{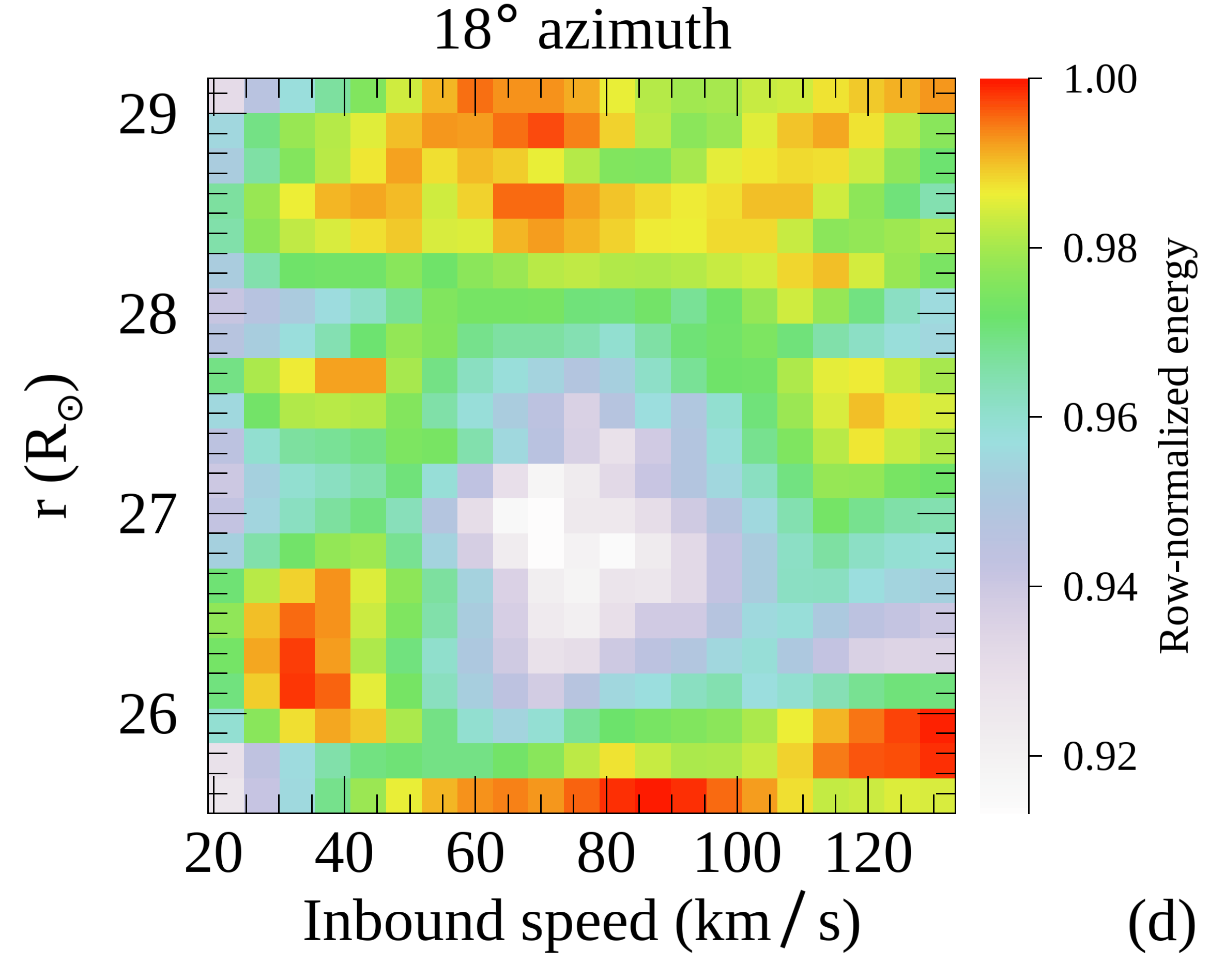}
    \caption{Inbound speed spectra as function of altitude, following Fig. \ref{fig:speed_spect}, for four selected azimuth. Panel (a) shows an example with a coherent ridge betwen \(\sim40\dash70\,\kmps\). Panel (b) shows an example with a coherent ridge around \(30\,\kmps\), which is close to the Fourier filter cutoff of \(19\,\kmps\). Panel (c) shows a case with a coherent ridge that may indicate acceleration from \(\sim50\,\kmps\) to \(\sim120\,\kmps\) between 29 and \(25\,\rs\). Panel (d) shows an example in which no coherent ridge is apparent, which is the case for the majority of azimuths.}
    \label{fig:speed_spect_reject}
\end{figure*}

To obtain a preliminary estimate of the prevalence of coherent inbound flows within the speed range expected near the outer edge of the Alfv\'en zone, we computed inbound speed spectra (as in the middle panel of Fig.~\ref{fig:speed_spect}) at azimuthal angles between 3\degree~and 357\degree, at 3\degree~cadence (3\degree,~6\degree, ..., 354\degree,~357\degree). The 119 speed spectra computed were visually inspected to estimate the number of azimuthal angles with clear inbound flows in the speed range of \(20\,\dash\,80\,\kmps\); for \(V_\text{sw}\) of \(300\,\kmps\) this corresponds to \(M_\text{A}\) between \(\sim 0.93\,\dash\,0.8\). The smallest row-normalized energy in the evaluated spectra was never greater than 0.9 and was often smaller than 0.8, giving us confidence in our distinction between the peak of the coherent ridge (normalized to unity) and the diffuse background in an individual spectrum. Fig. \ref{fig:speed_spect_reject}(a) shows another example of a spectrum with coherent inbound flow similar to the 344\degree~case in Fig. \ref{fig:speed_spect}. Fig. \ref{fig:speed_spect_reject}(b) shows a case with a coherent ridge around \(30\,\kmps\), which may be insufficiently distinct from the Fourier-filter cutoff at \(19\,\kmps\) to be considered physical. Fig.~\ref{fig:speed_spect_reject}(c) shows an example that may indicate acceleration with decreasing \(r\), which is consistent with the trend expected from a classical picture of the Alfv\'en speed getting larger with decreasing radial distance \citep{Cranmer2023SoPh}. Finally, Fig. \ref{fig:speed_spect_reject}(d) shows a case where no coherent speed ridge is apparent, which is the case for most of the azimuths. We identified 23 spectra with coherent inbound flow including types (a), (b), and (c), suggesting a filling fraction of \(\sim 0.19\) in azimuthal angle, between helioradii of \(\sim 26\) to \(29\,\rs\). 

The present estimate is not too different from the filling fraction of \(\sim0.1\) reported at these heights by \cite{Chhiber2024MNRAS} based on in-situ PSP observations in the ecliptic during the years 2021 to 2023. Besides the fact that the  specific time periods analyzed differ across the two studies, the larger estimate here may also be accounted for by two other factors: (i) 3D global modeling indicates that the Alfv\'en zone is shifted to greater heights outside the ecliptic, especially during low solar activity \citep{cohen2015SoPh,chhiber2019psp1,Chhiber2022MNRAS,Cranmer2023SoPh}; (ii) even within the ecliptic region sampled by PSP, the Alfv\'en zone shifts to greater helioradii during solar maximum, reaching median heights of \(15\,\dash\,23\,\rs\) \citep{Badman2025ApJ}. These factors are pertinent since our data are primarily comprised of measurements from outside the ecliptic, during an epoch of high solar activity.

\begin{figure}
    \centering
    \includegraphics[width=.35\textwidth]{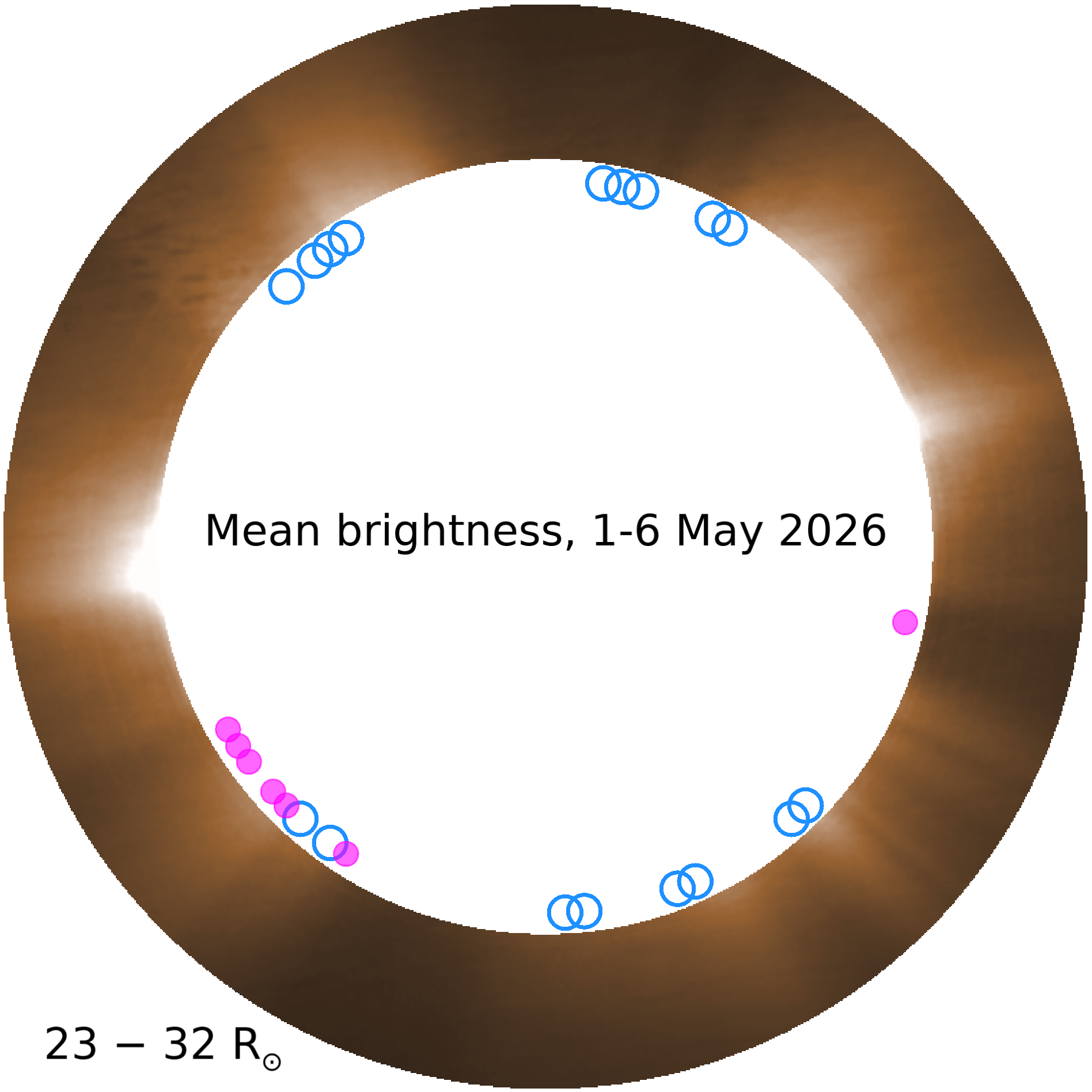}
    \caption{The image shows the mean brightness for each pixel during the indicated time period. Circles mark the azimuths where coherent inbound flow was detected, suggesting that the Alfv\'en zone extends to heights of \(23\,\dash\,32\,\rs\) at these locations. Open blue circles denote inbound speed spectra of the type shown in Fig. \ref{fig:speed_spect_reject}(a), while smaller, solid magenta circles represent spectra of the types in Fig. \ref{fig:speed_spect_reject}(b) and \ref{fig:speed_spect_reject}(c). See text for more details.}
    \label{fig:subAlf_azim}
\end{figure}

The azimuthal distribution of coherent inbound flow (in the speed range of interest) detected in our analysis is indicated in Fig. \ref{fig:subAlf_azim}, where the image is a representation of the mean brightness in the FoV during the time period of 1--6 May 2026; it shows the mean brightness of each pixel across all images within this period. With reference to Fig. \ref{fig:speed_spect_reject}, flows of type (a) are marked as open blue circles, while types (b) and (c) are marked as smaller, filled magenta circles. We surmise -- for the indicated azimuths, the upper edge of the Alfv\'en zone (or surface) lies within or above the range of heights examined here; the absence of coherent inbound flow in the remaining azimuths suggests that the Alfv\'en zone lies below \(23~R_\odot\) at these angles. In particular, low, equatorial solar latitudes appear to have lower Alfv\'en radii \citep[see also][]{chhiber2019psp1,Cranmer2023SoPh}. The pattern also suggests that the Alfv\'en zone is found at lower \(r\) for heliolatitudes associated with the brightest regions in the image, which may correspond to closed-loop streamers in the low corona; this would be consistent with results based on global MHD simulations \citep{cohen2015SoPh,chhiber2019psp1}.

%


%
\section{Conclusions and Discussion}
Despite the Alfv\'en surface's central importance in heliophysics \citep{Cranmer2023SoPh,Vidotto2021LRSP}, and in stellar astrophysics in general \citep{Matt2012ApJ,Chhiber2025ApJ}, its location and characteristics in three dimensions have not been well constrained by observations. In-situ measurements of the solar wind are usually limited to the ecliptic plane \citep[cf.][]{Verscharen2021MNRAS}, and inferences of Alfv\'en-surface/zone location tend to rely on assumed radial scalings of relevant parameters \citep[e.g.,][]{Badman2025ApJ}. One of the science objectives of the PUNCH mission is to map the global structure of the Alfv\'en zone via white-light imaging that covers a significantly expanded domain compared to in-situ datasets. The present study is a first step in this direction, employing preliminary science data from the PUNCH/WFI instrument to probe the outer edge of the Alfv\'en zone, between \(\sim23\,\dash\,32\,\rs\).

We employ an approach based on Fourier motion filtering to separate inbound flows (an indicator of sub-Alfv\'enic conditions) from the dominant outbound component within a temporal sequence of PUNCH images. Height-time images constructed from the filtered inbound dataset are interspersed with coherent downward ridges, suggestive of inbound flows at the outer edges of a patchy, fragmented Alfv\'en zone \citep{deforest2018ApJ,Chhiber2022MNRAS}. A speed spectrum analysis is performed to further distinguish coherent flows from noise, and to evaluate prevailing inbound flow speeds. This analysis suggests that coherent inbound flows with speeds between \(\sim30\,\dash\,80\,\kmps\) are present at a 0.19 fraction of azimuthal angles, within the range of helioradii examined. These speed values are consistent with expected Alfv\'en wave speeds in a sub-Alfv\'enic regime with \(M_A\sim0.9\). We surmise that the upper edges of the Alfv\'en zone extend to heights of \(23\,\dash\,32\,\rs\) within a small fraction of solar latitudes. This is consistent with a corrugated outer edge that occasionally protrudes to large altitudes, as has been noted in recent studies based on in-situ observations \citep{Verscharen2021MNRAS,wexler2021ApJ,Badman2025ApJ}.

We have made an attempt to be cautious in our interpretation, since the present analyses are based on preliminary PUNCH data that contain point-like artifacts and flicker; our pre-processing and Fourier filter are intended to address both issues. The preliminary data are also only available for \(r \gtrsim  23\,\rs\). These limitations of the early-mission data will be alleviated to a significant degree in the upcoming version 1 of PUNCH data. Accordingly, the present analysis will be refined in a future study that will include PUNCH observations of the entire range of heights spanned by the Alfv\'en zone, extending as low as \(5\,\rs\) by incorporating images from the Near Field Imager \citep[NFI;][]{Colaninno2025SoPh}.

We conclude with two important caveats that underlie any approach toward mapping the Alfv\'en zone via white-light imaging. First, the brightness features measured by such instruments correspond to density inhomogeneities, so the observations do not directly `see' incompressible Alfv\'en waves \citep{Cranmer2023SoPh}. However, the observed density variations may arise from propagating magnetosonic waves, and the fast magnetosonic speed is expected to be quite close to \(V_A\) in the corona. Mode conversion from the magnetosonic to Alfv\'en mode can also produce a direct relationship between density variations and the usual Alfv\'enic oscillations \citep{Morton2023RvMPP}. Myriad solar-wind phenomena like shocks, jets, and reconnection exhausts can produce observable density perturbations that flow at speeds of order \(V_A\) as well \citep{Gosling2006JGRA,Warmuth2015LRSP,Cranmer2021ApJ,Raouafi2023ApJ}. In nearly all cases, inbound propagation would indicate a sub-Alfv\'enic medium. 

Second, line-of-sight integration through the optically thin solar wind plasma can superpose structures at different distances and projected speeds, broadening the computed speed spectrum. The inferred speeds therefore describe projected brightness patterns, and cannot, without additional geometric assumptions, be identified uniquely with localized plasma or wave motions. Note that non-radial motions along curved coronal loops are less relevant at the analyzed distances of \(23\,\dash\,32\,R_\odot\), where solar-wind flow is predominantly radial. Future studies may employ tomographic techniques \citep{Alzate2023ApJ} and validation approaches using synthetic or simulation-based data \citep{Pecora2024SoPh,MoraesFilho2024ApJ,Cranmer2026ApJ} to attempt disambiguation. Further consideration of these concerns is deferred to future studies with non-preliminary PUNCH data. 


%
\begin{acknowledgments}
PUNCH is a heliophysics mission to study the corona, solar wind, and space weather as an integrated system, and is part of NASA's Explorers program (Contract 80GSFC14C0014).
This work was supported at the University of Delaware by the PUNCH mission through subcontract N99054DS from NASA/SWRI. It was also partially supported by the NASA Living With a Star (LWS) Science program grant 80NSSC22K1020. The National Center for Atmospheric Research is a major facility sponsored by the National Science Foundation under Cooperative Agreement No. 1852977.
\end{acknowledgments}
\appendix
\section{A Generalized 2D Hampel Filter for Automated Detection and Correction of Point-like Artifacts in Images}\label{sec:Hampel}
The standard median filter (SMF) employs a sliding-window approach in which the locally-computed median is used for smoothing 1D data \citep{Tukey1977ExplorData}. Its strength lies in its strong resistance to local outliers or impulsive noise, but its reliance on a single tuning parameter (the window length) limits its flexibility, and the filtering operation can introduce significant distortion in the portion of the signal that we do not wish to discard. The \textit{Hampel filter} is an extension of the SMF that replaces the central value in the data window with the median \textit{only} if it lies sufficiently far from the median to be deemed an outlier \citep{Davies1993JASA_Hampel}. Importantly, the measure of deviation is the \textit{median absolute deviation} (MAD) rather than the standard deviation (\(\sigma\)). This makes the Hampel filter distinct from the \(\sigma\)-clipping approach widely employed in astronomy \citep{Masias2012MNRAS}, which typically makes decisions based on mean/standard-deviation computations that are themselves influenced by the presence of outliers.

The Hampel filter has recently found fruitful application in heliophysics to address ``spiky'' noise in time series, both in the temporal and spectral domains \citep{bandyopadhyay2018filter,chhiber2020ApJS}. In this study, we employ a 2D generalization of the Hampel filter to detect and correct point-like\footnote{We use the term `point-like' to describe the visual appearance of the artifacts in the context of the entire image, while bearing in mind that they may occupy several pixels.} artifacts in the preliminary PUNCH images; these artifacts are very-local enhancements or depletions in the brightness (see Fig. \ref{fig:Hampel}, top panel) that are related to starfield suppression and other processing steps involved in producing v0l L3 data. We describe our implementation below.

Let \(x(\bm{p})\) denote the image value at pixel \(\bm{p}\), and let \(W_{\bm{p}}\) be a local \(w\times w\) neighborhood centered on that pixel. We first compute the local median
\begin{equation}
    m(\bm{p}) = \text{median}_{\bm{q}\in W_{\bm{p}} }~x(\bm{q}).
\end{equation}
The absolute residual from this local median is
\begin{equation}
    d(\bm{p}) = |x(\bm{p}) - m(\bm{p})|,
    \label{eq:Hampel_residual}
\end{equation}
and a local measure of dispersion relative to the median is defined as\footnote{The factor \(1.4826\) makes this estimate consistent with an estimate of the standard deviation for Gaussian data \citep{Pearson2016EJASP}; it is not relevant to the practical implementation since the \(n_\text{tol}\) parameter, in effect, further multiplies \(s(\bm{p})\) with an arbitrary numerical factor.}
\begin{equation}
    s(\bm{p}) = 1.4826~\text{median}_{\bm{q}\in W_{\bm{p}} } d(\bm{q}).
\end{equation}
A pixel is identified as an outlier when \(d(\bm{p})>n_\text{tol}~s(\bm{p})\), and flagged pixels are replaced by the corresponding local median:
\begin{equation}
x_{\rm filt}(\bm{p}) =
\begin{cases}
m(\bm{p}), &
d(\bm{p})
> n_{\rm tol}\,s(\bm{p}), \\[4pt]
x(\bm{p}), &
\text{otherwise}.
\end{cases}
\end{equation}

Note that there is a subtle distinction between the median deviation \(s(\bm{p})\) as computed above and the MAD typically employed in the Hampel filter:
\begin{equation}
    \text{MAD}(\bm{p})= 1.4826~\text{median}_{\bm{q}\in W_{\bm{p}}} 
          |x(\bm{q})-m(\bm{p})|;
\end{equation}
i.e., every value in the window is compared with the \textit{same} central-pixel window median \(m(\bm{p})\). Our implementation employs Eq. \eqref{eq:Hampel_residual} instead since this operation is easily vectorized and allows one to avoid loops, which are computationally inefficient in some high-level programming environments. While this approach was developed independently during the present analysis, it recently came to our attention that \cite{Toribio2025Sensors} have proposed a similar implementation in the context of 1D timeseries; they also evaluate the statistical robustness of the Hampel-filter variant based on this modified-MAD (mMAD) approach.
\begin{figure}
    \centering
       \includegraphics[width=.4\textwidth]{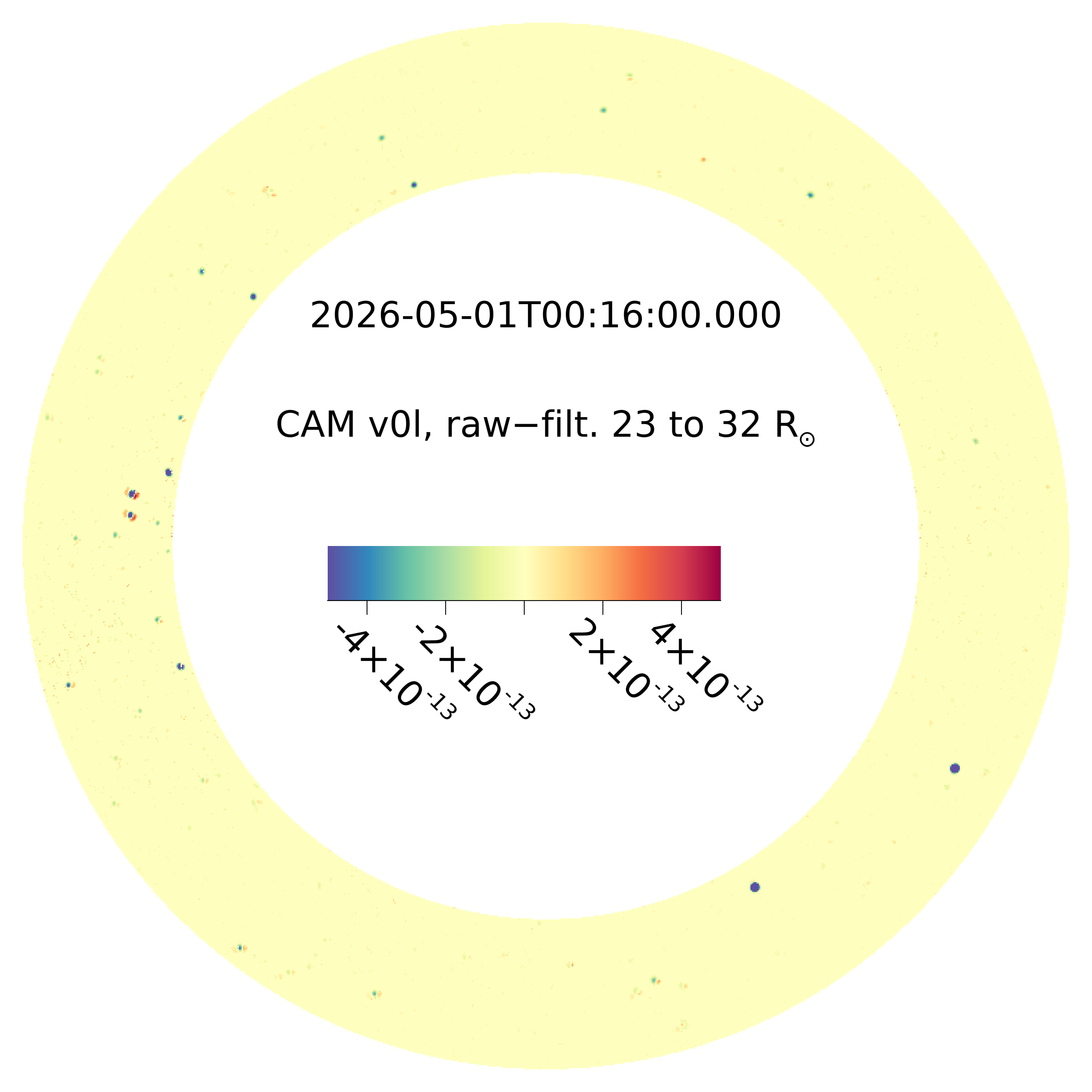}
    \caption{The residual image produced by forming the difference between the raw and Hampel-filtered images shown in Fig.~\ref{fig:Hampel}. The filter identifies point-like outliers without disturbing the vast majority of pixels.}
    \label{fig:Hampel2}
\end{figure}

We set \(w=19\) pixels, and \(n_\text{tol}=3\); these parameters work reasonably well for our purpose in this study; the bottom panel of Fig. \ref{fig:Hampel} shows an example filtered image. The residual image formed by taking the difference between the raw and filtered images (i.e., the top and bottom images in Fig. \ref{fig:Hampel}) is shown in Fig. \ref{fig:Hampel2}. Evidently, the filter is effective at identifying point-like artifacts without disturbing the vast majority of pixels (cream-colored regions in Fig. \ref{fig:Hampel2}).

Our parameter choices are obviously not universal. Future work can attempt a more systematic parameter study, and also implement multi-level `cascades' or recursive filtering \citep{Pearson2016EJASP} to correct artifacts associated with different spatial scales.


\end{document}